\documentclass[3p,11pt]{elsarticle}
\usepackage{graphicx}
\usepackage{lineno}
\usepackage{hyperref}
\usepackage{color}
\usepackage[figuresleft]{rotating}
\usepackage{caption}
\usepackage{subcaption}
\usepackage{layout}
\usepackage{setspace}
\usepackage{dcolumn}
\usepackage{xtab}
\usepackage[T1]{fontenc} 
\usepackage{tgtermes, tgheros} 
\usepackage[scaled=0.92]{helvet}
\usepackage{autonum}
\usepackage{mathtools}
\usepackage{multirow}
\usepackage{amsfonts}
\usepackage{bm}

\newcommand{\Var}{\mathrm{Var}}
\newcommand{\Cov}{\mathrm{Cov}}

\usepackage{array}
\usepackage{tabularx}
\newcolumntype{C}{>{\centering\arraybackslash}X}
\newcolumntype{L}{>{\raggedright\arraybackslash}X}
\newcolumntype{R}{>{\raggedleft\arraybackslash}X}
\usepackage[table]{xcolor}
\newcommand*{\diff}{\mathrm{d}}
\journal{AAA}
\biboptions{authoryear}
\usepackage{hyperref}
\makeatletter
\def\ps@pprintTitle{%
\let\@oddhead\@empty
\let\@evenhead\@empty
\def\@oddfoot{}%
\let\@evenfoot\@oddfoot}
\makeatother
\begin{document}
\onehalfspacing
\begin{frontmatter}
\title{
Restoration and extrapolation of structural transformation
\\by dynamical general equilibrium feedbacks  
}
\author[sn]{Satoshi Nakano}
\ead{nakano@jil.go.jp}
\author[kn]{Kazuhiko Nishimura
}
\ead{nishimura@n-fukushi.ac.jp}
\address[sn]{The Japan Institute for Labour Policy and Training, Nerima Tokyo 177-0044, Japan}
\address[kn]{Faculty of Economics, Nihon Fukushi University, Tokai Aichi 477-0031, Japan}
\begin{abstract}
We model sectoral production by serially nesting binary compounding processes.
The sequence of processes is discovered in a self-similar hierarchical structure stylized in macroscopic input-output transactions.
The feedback system of unit cost functions, with recursively estimated nest-wise CES parameters, is calibrated for sectoral productivities to replicate two temporally distant cost share structures, observed in a set of linked input--output tables.
We model representative households by multifactor CES, with parameters estimated by fixed effects regressions.
By the integrated dynamical general equilibrium model, we extrapolate potential structural transformations, and measure the associated welfare changes, caused by exogenous sectoral productivity shocks.
\end{abstract}
\end{frontmatter}

\section{Introduction}
Structural transformation refers to the reallocation of economic activity across a broad range of industrial sectors.\footnote{\cite{KM} and \citet{BH6} present seminal reviews of structural change/transformation.}
It has played an important role in the growth of developing economies.
Based on macroeconomic dynamic optimization frameworks, previous studies have focused on understanding stylized facts of structural transformation among and across different regions.
Structural transformation has been studied in relation to sectoral productivity, capital intensity, international trade, scale economy, and technologies embodied in the elasticity of factor substitution, among others.\footnote{Important contributions, besides the classics, include \citet{ngaietal, eche, acegue, buera, herrenAER, a-c}, among others.}
While these models explain long-term transitions of agrregate economic indicators, the underlying framework must be applicable for studying innovative activities, perhaps of the current or short-term economy, regarding a broader range of sectoral transactions.

This study extends the framework of structural transformation in this direction.
In particular, we examine structural transformations concerning a wide range of sectoral activities comprising a wide range of factor inputs.
Accordingly, we modify the typical two-factor sectoral production functions employed in previous models. 
We consider multiple ($N=385$) sectoral production functions with multiple (potentially 385) intermediate factor inputs, regarding sector and commodity classifications of the Japanese input--output tables, in addition to the two primary factor inputs, i.e., labor and capital services.
For our purposes, structural transformation is not limited to a possible shifting of shares between capital and labor.
Instead, we examine observed and potential shifts in the sectoral cost share structures of all factor inputs, regarding potential factor substitutions.

This study mainly aims to construct an empirical macroeconomic general equilibrium model concerning many commodities that are produced, consumed, and used as inputs of production.
The pioneering contribution to the economy-wide modeling of production is Leontief's input--output system.
Even now, in the applied (computable) general equilibrium modeling literature, input--output tables and constant returns to scale production function play a key role, especially for intermediate productions \citep[][]{burfisher}.
In contrast, we create another variant of production function that can endogeize transformations of factor cost share structures (or input coefficients) without imposing uniform (zero or unit or any) a priori elasticity of factor substitutions.

Previous applied general equilibrium analyses can be classified by production function characteristics.
In the computable general equilibrium literature, constant elasticity of substitution (CES) functions have been extensively applied.
Two-factor CES functions \citep{acms} have been modified for empirical purposes in two broad aspects: one to relax the constant returns assumption, and the other to relax the two-factor restrictions.
Estimating the elasticity of substitution of a non-constant returns to scale CES function involves non-linear regression.
\citet{kmenta} used a Taylor series approximation to linearize the regression.
Regarding multiple factor inputs, \citet{sato} nested CES functions in two stages. 
A CES function allows only single substitution elasticity among multiple inputs \citep{uzawa, mcfadden}.

Apart from CES, \citet{hudson} developed empirical translog functions with four aggregated factors, which became the precursor to subsequent KLEM-type econometric general equilibrium models.
A second-order generalization of Cobb--Douglas, translog functions are flexible with regard to substitution elasticities across factor inputs.
For their empirical estimation, cost (expenditure) shares are typically utilized using a dual approach, in addition to first-order conditions \citep{handbook}.
While first order conditions have been central in estimating substitution elasticities and productivity growths for two-factor CES functions from time series observations \citep[e.g.,][]{berndt, antras, normces, herrenAE}, \cite{knn} took a dual approach in estimating CES functions with multiple factors from cost share data observed in linked input--output tables.

We also model sectoral production, nevertheless, by serially nesting (i.e., cascading) two-factor CES functions, to allow for multiple factor inputs. 
We show that the nest-wise elasticity and share parameters of a cascaded CES function can be estimated with a constrained least squares minimization problem solvable via dynamic programming.
We base our empirical study on linked input--output tables, where monetary transactions between all sectors and commodities are available in nominal and real values for different periods.
The parameters of our empirical model are estimated at two points (i.e., a two-point regression), from which the parameters become state-restoring.
Under our set of empirical cascaded CES functions, the observed input coefficients of the two periods are completely restored.
Furthermore, sectoral total factor productivity (TFP) growth measured under these functions, reveals a high level of concordance with those (i.e., T\"{o}rnqvist) that are consistent with the underlying translog functions.

Cost share data (or, input coefficients) in this study are drawn from the linked input--output tables \citep{miac}, with the following two-way balances:\footnote{Our two-period linked input--output tables are basically created by averaging two neighboring pairs of tables from three years (2000, 2005, 2011).
Thus, we have a set of two tables, with $X_{ij1} >0$ only if $X_{ij0} >0$, and vice-versa; in other words, the variety of factor inputs is preserved between the two periods for each sector. }
\begin{align}
r_{t} K_{jt} + w_{t} L_{jt} + \sum_{i=1}^N p_{it} X_{ijt}  = p_{jt} {Y}_{jt} &&
H_{it} + G_{it} + E_{it} + \sum_{j=1}^N X_{ijt} = {Y}_{it} 
\label{liot}
\end{align}
All symbols represent available data.
Here, $i=1,\cdots,N$ indicates a commodity, $j=1,\cdots,N$ a sector, and $t=0,1$ a period concerned.
The commodity price $(p_{1t}, \cdots, p_{Nt})$ are observable for the two periods $t=0,1$ since linked input--output tables are recorded in both nominal and real values.
Capital service price $r$ and averaged wage rate $w$ are drawn from \citet{jip} database.
$X$ and $Y$ represent intermediate factor inputs and commodity outputs, respectively.
Regarding the left-hand side identity, $K$ and $L $ represent capital service and labor inputs, respectively.
Regarding the right-hand side identity, $H$, $G$, and $E$ represent household consumption, fixed capital formation, and net exports, respectively.
We note by construction that the value added $\sum_{j=1}^N ( r_t K_{jt} + w_t L_{jt})$ equals final demand $\sum_{i=1}^N p_{it} (H_{it} + G_{it} + E_{it})$.

The general equilibrium feedback system of dual (unit cost) functions that we derive traces the observed structural transformation under the fundamental (binary compounding) structure of production, which we assume to be immutable.
In other words, for modeling sectoral activities in terms of binary compounding processes, a persistent sequence of processes (nests) must be specified.
In this regard we use the sequence of sectors latent in input--output tables.
If production comprises only binary compounding processes, factor inputs compile as the process compounds.
When each compound is viewed as an intermediate output, it compiles as the process compounds further, after which the input--output structure becomes triangular at all levels of production.
This scale-freeness in a triangular structure is utilized to specify the intrasectoral sequence of compounding processes.

The remainder of the paper is organized as follows.
The next section specifies the nesting order (i.e., a stream order) of binary compounding processes applicable in all sectoral productions by triangulating the input--output incidence matrix.
Section 3 indicates how the parameters of a cascaded CES function can be estimated via dynamic programming and observes that structural transformation can fully be traced by the restoring parameters.
Section 4 estimates the indirect utility function of a representative household with multifactor CES by using a fixed effects regression.
Section 5 integrates sectoral production with representative household consumption and demonstrates dynamic general equilibrium feedbacks with endogenized structural transformations.
Section 6 provides concluding remarks.

\section{Fundamental Structure}
Consider a cascaded production comprising ${N}$ binary processes $(j= 1, \cdots, {N})$ compounding ${N}+1$ inputs $(i=0,\cdots,{N})$ in an ascending order of index $i$.
Define incidence such that $\phi_{ij} =1$ if input $i$ enters process $j$ directly or indirectly, and $\phi_{ij}=0$ if input $i$ never enters process $j$ even indirectly.
For the case of a cascaded production, the incidence matrix ${\Phi} =\left\{ \phi_{ij} \right\}$ becomes triangular, i.e., $\phi_{ij}=1$ iff $i \leq j$ and $\phi_{ij}=0$ iff $i > j$.
Furthermore, 
every process $j=1,\cdots,{N}$ constitutes part of an entire sequence of the compounding processes. 
That is, intermediate product $j$ (or the output of the $j$th compounding process) is produced by compounding inputs $i=0,\cdots,j-1$, in this order, which constitutes the first $j$ part of the entire sequence. 
Given that the underlying production is binary compounding, the processing sequence is unraveled if the ordering of inputs (or binary processes) makes the incidence matrix triangular.\footnote{To this end, however, any circular flow must be ruled out \citep[][]{nn}.}

Let us now focus on the $k$th process of an ${N}$ cascading production.
Define $\sum_{i=1}^{N} \phi_{ik}$ and $\sum_{j=1}^{N} \phi_{kj}$ as \textit{indegree} and \textit{outdegree} of the $k$th process, respectively.
For a perfectly triangular incidence matrix $\Phi$, the indegree-outdegree ratio of the $k$th process will be evaluated as follows:\footnote{\citet{cw} used the same criteria (ratios between indegree and outdegree) for categorizing industrial sectors, except that they used input coefficients $a_{ij}$ instead of incidents $\phi_{ij}$. 
For similar purposes \citet{antrasGVC} applied the concept of average propagation length.
}
\begin{align}
\text{Indegree/outdegree of $k$}
= \frac{\sum_{i=1}^{N} \phi_{ik}}{\sum_{j=1}^{N} \phi_{kj}}
= \frac{k}{{N}-k+1} ~\text{ (for a perfectly triangular $\Phi$)}
\end{align}
Besides that, it is convenient to use the following \textit{ranking index} to indicate the $k$th rank out of $N$ alternatives:
\begin{align}
\text{Ranking index of $k$}
= \frac{{N}-k+1}{{N}}
\end{align}
Sorting ${N}$ observed values in ascending order and plotting against the ranking index gives the complementary cumulative density function (CCDF) of the observed values (with equal probability).

\begin{figure}[t!]
\centering
\begin{minipage}[b]{0.475\textwidth}
\includegraphics[width=\textwidth]{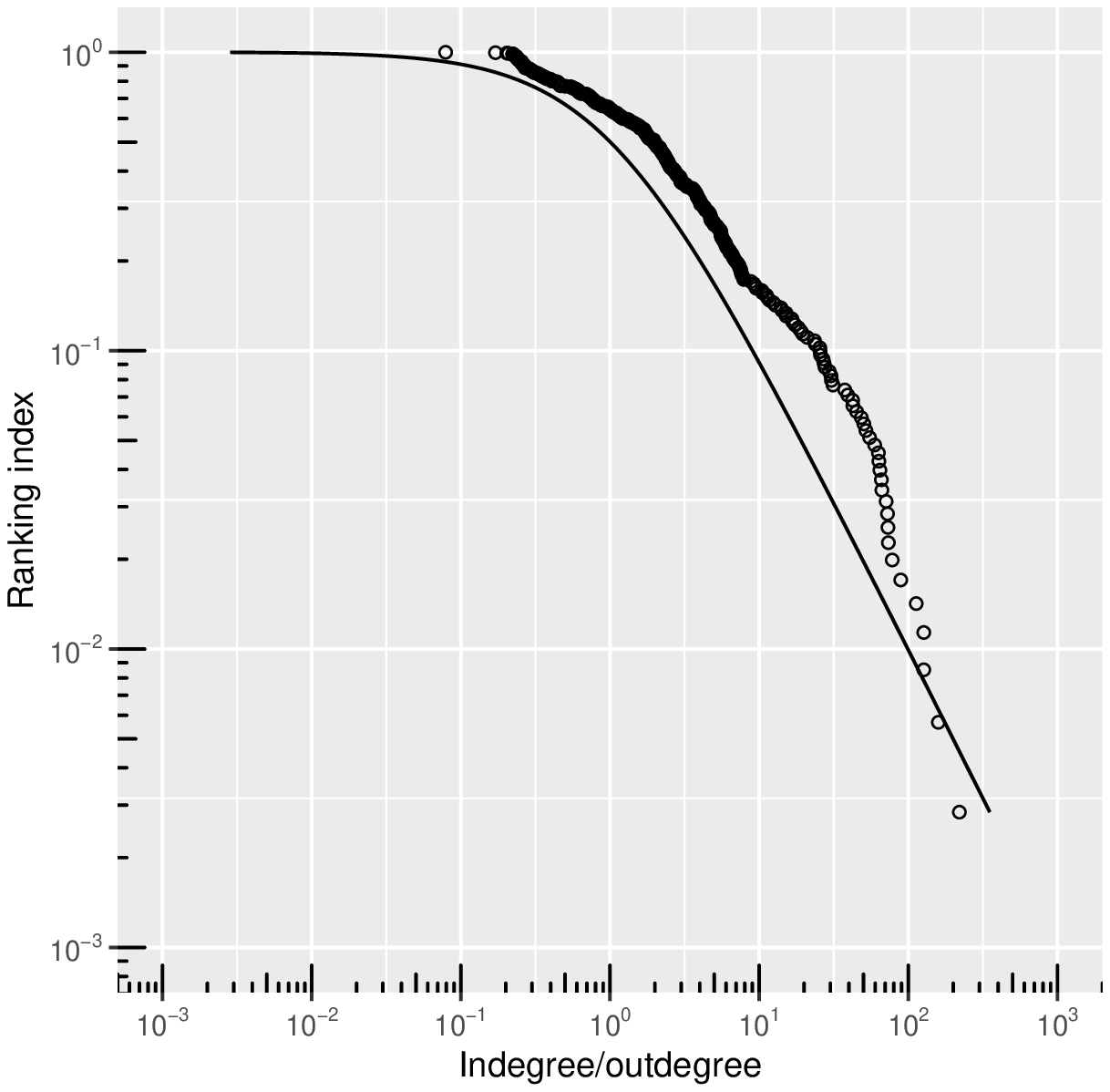}
\caption{Open dots correspond to the CCDF of indegree/outdegree values of the input--output incidence matrix of Japan 2005. The solid line is the CCDF of indegree/outdegree values of a perfectly triangular incidence matrix.  }
\label{fig_power}
\end{minipage}
\hspace{0.03\textwidth}
\begin{minipage}[b]{0.475\textwidth}
\includegraphics[width=\textwidth]{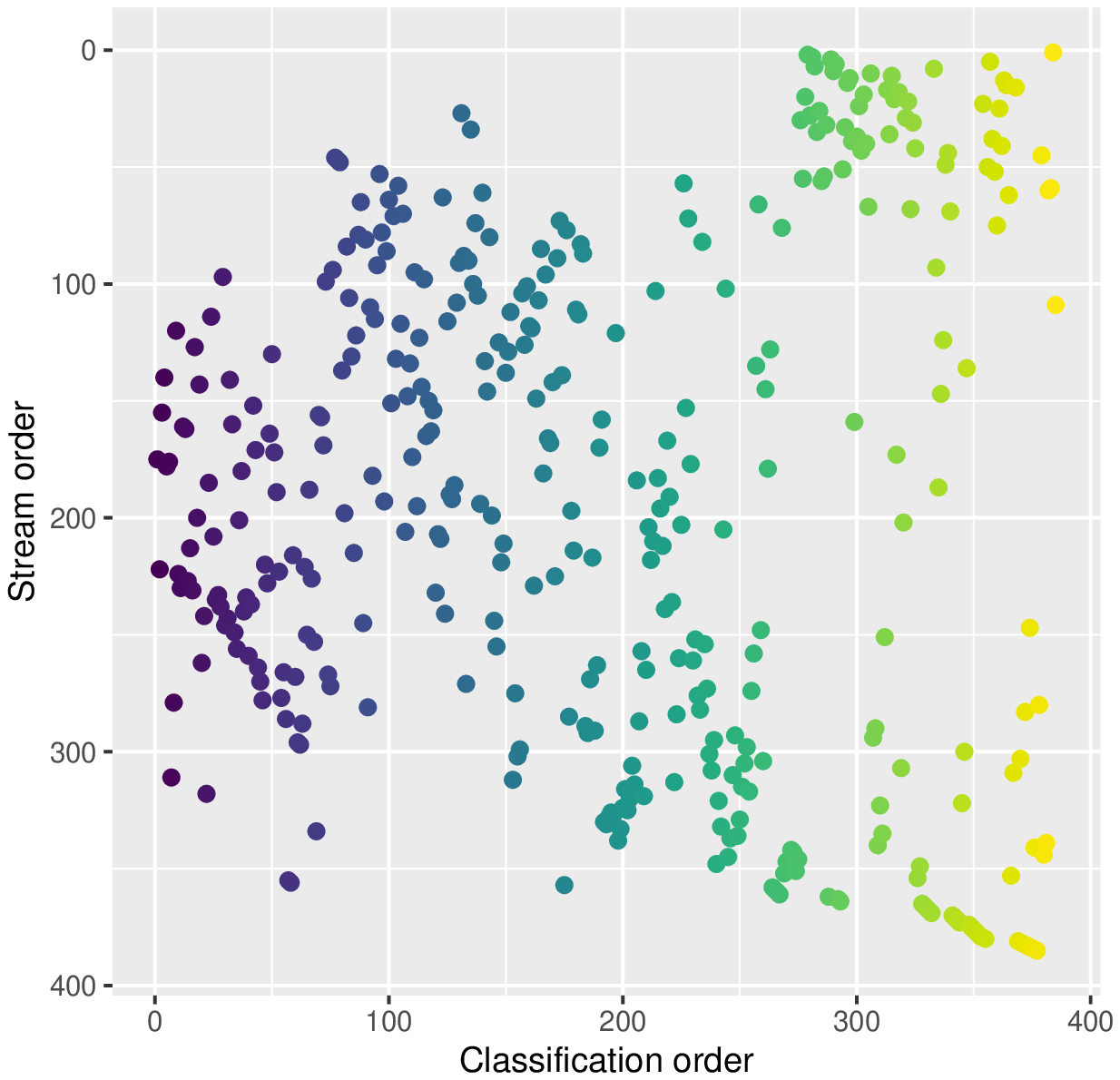}
\caption{Low indegree/outdegree values correspond to upstream (at the top) of a stream order.  Classification order is based on Colin Clark's primary (1--32), secondary (33--263), and tertiary (264--385) classifications.}
\label{fig_class}
\end{minipage}
\end{figure}

In Figure \ref{fig_power}, we plot, in a solid line, the ranking index of indegree/outdegree values of an incidence matrix $\Phi$ representing a cascading production (which should be perfectly triangular). 
In this case, the indegree/outdegree values of the $k$th process must be ranked $k$th in the ranking index.
We may observe linearity between the log of the two functions as $k$ approaches $N$, i.e., $\log \frac{{N}-k+1}{{N}} \approx - \log \frac{k}{{N}-k+1}$, indicating asymptotic power-law relationships between them.\footnote{In many cases power-law distribution implies scale freeness and self-similarity \citep{ecolett}.}
In the same figure we also plot, in open dots, 
the indegree/outdegree values of the incidence matrix created from the 2005 input--output table of Japan (with $\phi_{ij} = 1$ iff $X_{ij}>0$ and $\phi_{ij}=0$ otherwise) in an ascending order against the corresponding ranking index.\footnote{Note that we designate primary factors (or value added) for $i=0$ so that $\phi_{0j} = 1$ for all $j$.}
If the $\mathcal{N}$ processes spanning the entire economy were aggregated into $N$ sectors without spoiling the hierarchy of processes, the input--output table of $N$ sectors must also be triangular, and its ranking index should represent the economy-wide sectoral processing from upstream to downstream. 
For empirical purposes, we apply this hierarchy, which we hereafter call the \textit{stream order}, in all sectoral productions as well.
Figure \ref{fig_class} shows correspondences between stream order and input--output table's \textit{classification order} which is based on Colin Clark's three-sector model.

\section{Sectoral Production }
\subsection{Two-Factor CES Production}
We start with the following two-factor CES production function of constant returns to scale:
\begin{align}
q
= f\left(K, L \right)
=
\left( \alpha^{\frac{1}{1-\gamma}} K^{\frac{\gamma}{\gamma-1}} + (1-\alpha)^{\frac{1}{1-\gamma}} L^{\frac{\gamma}{\gamma-1}} \right)^{\frac{\gamma -1}{\gamma}}
\label{2ces}
\end{align}
where $1-\gamma$ denotes the elasticity of substitution.
$K$ and $L$ are the first and second factor inputs in physical quantity, respectively.
$q$ denotes the output, in physical quantity. 
The share parameter $0 \leq \alpha \leq 1$ and $\gamma$ are assumed to be constant over time.
For convenience, we work on the following dual function of (\ref{2ces}):
\begin{align}
\pi 
=c\left( r, w \right)
=\left( \alpha r^{\gamma} + (1-\alpha) w^{\gamma} \right)^{\frac{1}{\gamma}}
\label{2cesuc}
\end{align}
where $r$ and $w$ denote factor prices for the first and second inputs, respectively.
The unit cost of $q$ is denoted by $\pi$.
Duality between (\ref{2ces}) and (\ref{2cesuc}) implies $\pi q = rK + wL$, or, 
\begin{align}
\pi
=r K/q + wL/q
= r\frac{\partial \pi}{\partial r} + w\frac{\partial \pi}{\partial w} 
\label{cost}
\end{align}
The second identity of (\ref{cost}) is due to Euler's theorem applied to (\ref{2cesuc}), which is homogeneous of degree one. 

Cost shares, which we denote by $a$ for the first factor and $1-a$ for the second factor, can be restated by the following expansion with regard to (\ref{2cesuc}) and (\ref{cost}):
\begin{align}
a = \frac{rK}{\pi q}
=\frac{r}{\pi}\frac{\partial \pi}{\partial r}
= \alpha \left( \frac{r}{\pi} \right)^\gamma
&&
1-a = \frac{wL}{\pi q}
=\frac{w}{\pi}\frac{\partial \pi}{\partial w}
= (1-\alpha) \left( \frac{w}{\pi} \right)^\gamma
\end{align}
Then, we can work on the ratio of the two cost shares as follows:
\begin{align}
\frac{a}{1-a} 
= \frac{rK}{wL}
= \frac{\alpha}{1-\alpha} \left( \frac{r}{w} \right)^\gamma
\label{ratio}
\end{align}
By taking the natural log and indexing samples by $t$, we have a linear regression equation as follows:
\footnote{Although essentially equivalent, \citet{acms}, \citet{berndt}, and \citet{antras} use $K/L$ in the regressand.
The CES functional form was derived from the regression equation by \citet{acms}. 
As we measure all prices relative to those of the latest period, the parameters are normalized \citep{KdLG, normces} at the latest period. }
\begin{align}
\ln {{z}}_t = \ln \zeta + \gamma \ln r_t/w_t + \varepsilon_t
\label{regtwo}
\end{align}
where, we define that ${z}=\frac{a}{1-a} = \frac{rK}{wL}$ and $\zeta=\frac{\alpha}{1-\alpha}$. 
The error terms are denoted by $\varepsilon$.
Below we present the OLS estimators for the unknown parameters:
\begin{align}
&\hat{\gamma} = \frac{\Cov\left(\ln {{z}_t}, \ln {r}_t/{w}_t\right)}{\Var\left( \ln {r}_t/{w}_t\right)}
&\ln \hat{\zeta} 
= \frac{\sum_{t=0}^T\ln {{z}_t} - \hat{\gamma} \sum_{t=0}^T{\ln r_t/w_t}}{T+1}
\label{parmhat}
\end{align}

\subsection{Restoring Parameters}
With a time dimension of two i.e., $t=0,1$, 
(\ref{regtwo}) becomes a two-point regression (or, a before-after analysis), in which the parameters are solved with the following solution:
\begin{align}
\bar{\gamma} = \frac{\ln {{z}}_1 -\ln {{z}}_0}{\ln r_1/w_1 - \ln r_0/w_0}
&&
\ln \bar{\zeta} = \frac{\ln {{z}}_0 \ln r_1/w_1 -\ln {{z}}_1 \ln r_0/w_0}{\ln r_1/w_1 - \ln r_0/w_0} 
\label{parm2p}
\end{align}
Here, we indicate the two-point estimator by a bar instead of a hat.
In this case, the error terms are null, i.e., $\varepsilon_0=\varepsilon_1=0$.
In other words, these parameters are \textit{state-restoring} in the sense that two observed cost shares are completely restored by the two observed factor price ratios via (\ref{ratio}) such that
\begin{align}
\frac{a_0}{1-a_0}= \frac{\bar{\alpha}}{1-\bar{\alpha}} \left( \frac{r_0}{w_0} \right)^{\bar{\gamma}}
&&
\frac{a_1}{1-a_1}= \frac{\bar{\alpha}}{1-\bar{\alpha}} \left( \frac{r_1}{w_1} \right)^{\bar{\gamma}}
\end{align}
where $\bar{\alpha}={\bar{\zeta}}/{(1+\bar{\zeta})}$ and ${a}_t={{z}_t}/{(1+{z}_t)}$, as defined previously.
Note that $\bar{\alpha}=a_t$ if the parameters are normalized at $t$.
Moreover, the unit cost $\pi_t$ (of output $q_t$) can be evaluated via (\ref{2cesuc}) as follows:
\begin{align}
\pi_0 = \left( \bar{\alpha} (r_0)^{\bar{\gamma}} + \left( 1 - \bar{\alpha} \right) (w_0)^{\bar{\gamma}} \right)^{\frac{1}{\bar{\gamma}}}
&&
\pi_1 = \left( \bar{\alpha} (r_1)^{\bar{\gamma}} + \left( 1 - \bar{\alpha} \right) (w_1)^{\bar{\gamma}} \right)^{\frac{1}{\bar{\gamma}}}
\end{align}

\subsection{Cascaded CES Production}
A cascaded (or serially nested) compounding function of $N+2$ inputs (two primary and $N$ intermediate inputs) for an industrial sector (whose index $j$ is omitted) can be described as follows:
\begin{align}
Q &=  F \left( K, L, X_1, \cdots, X_{N} \right) 
\notag \\
&= f_{N}\left( X_{N}, {f}_{N-1} \left( X_{N-1}, \cdots f_2 \left( X_2, {f}_1\left( X_1, f_0\left( K, L \right) \right) \right) \cdots \right) \right)  
\end{align}
Here, $Q$ and $X_i$ denote the output of production and the $i$th input, respectively. 
There are $N+1$ nests, each comprising a single input and a compound from the lower-level nest, except for the primary nest, which includes two factor inputs, $K$ and $L$. 
We let $q_1=f_0 \left( K, L \right) $ be the 0th (aggregated primary) factor input. 

Assume that the $n+1$th compound output of the $n$th nest is given by the CES production function.
\begin{align}
{q}_{n+1} = f_{n}\left( X_{n}, {q}_{n} \right) = \left( ({\alpha}_{n})^\frac{1}{1-\gamma_{n}} (X_{n})^\frac{\gamma_{n}}{\gamma_{n}-1} + (1-{\alpha}_{n})^\frac{1}{1-\gamma_{n}} ({q}_{n})^\frac{\gamma_{n}}{\gamma_{n}-1} \right)^\frac{\gamma_{n}-1}{\gamma_{n}}
\label{npf}
\end{align}
The nest production function (\ref{npf}) applies for $n =0, 1, \cdots, N$, and we suppose that $X_0 = K$ and $q_0 = L$.
Here, $0\leq {\alpha}_n \leq 1$ denotes the share parameter of the $n$th nest,
${q}_{n}$ is the compound output from the $n -1$th nest, and 
$1-\gamma_{n}$ is the elasticity of substitution between $X_{n}$ and ${q}_{n}$.
We assume that (\ref{npf}) exhibits constant returns to scale.
The following is the dual of the production function given in (\ref{npf}):
\begin{align}
\pi_{n+1} = c_{n}\left( p_{n}, \pi_{n} \right) = \left( {\alpha}_{n} (p_{n})^{\gamma_{n}} + (1-\alpha_{n}) (\pi_{n})^{\gamma_{n}} \right)^\frac{1}{\gamma_{n}}
\label{ncf}
\end{align}
Here, $p_n$ denotes the price of the $n$th input, and $\pi_{n}$ denotes the unit cost of the compound from the lower level nest.
Since (\ref{ncf}) must also hold for $n=0$, we let $p_0 = r$ and $\pi_0 = w$ for consistency with (\ref{npf}).
A cascaded CES dual function can be created by serially nesting (\ref{ncf}) as follows:
\begin{align}
{\Pi} &= C \left( r, w, p_1, \cdots, p_{N} \right) 
\label{ccesucf}
\\
&= c_{N}\left( p_N, c_{N-1} \left( p_{N-1}, \cdots c_2 \left( p_2, c_1\left( p_1, c_0 \left( r, w \right) \right) \right) \cdots \right)\right) 
\end{align}
where $\Pi=\pi_{N+1}$ denotes the unit cost of the output $Q$ for the sector concerned.

\subsection{Parameter Estimation}
Due to zero profit condition (of constant returns to scale) in all nests, below must hold:
\begin{align}
\pi_{n+1} q_{n+1} = p_n X_n + \pi_{n} q_{n}
\label{zerop}
\end{align}
By applying Shephard's lemma to the unit cost function (\ref{ccesucf}) and considering (\ref{zerop}) gives the following: 
\begin{align}
\frac{p_n}{\Pi}\frac{\partial C }{\partial p_n} = 
\frac{p_n X_n}{\Pi Q}
= a_n
&&
\frac{\pi_n}{\Pi}\frac{\partial C }{\partial \pi_n} 
=
\frac{\pi_n q_n}{\Pi Q}= \sum_{i=0}^{n-1} a_i
\label{aaa}
\end{align}
where $a_n$ denotes the cost share of the $n$th input.
Cost shares are observable from our data (\ref{liot}) by $a_n = \frac{p_n X_n}{p Y}$. 
Note that $\Pi Q = p Y$ holds due to zero profit condition.
Also, note that (\ref{aaa}) stand for $n=0, \cdots, N$ as we allow
$\sum_{i=0}^{-1}a_i = a_0 -a_{-1} = \frac{\pi_0 q_0}{\Pi Q} = \frac{w L}{p Y}=a_L$, and
$a_0 =\frac{p_0 X_0}{\Pi Q} = \frac{r K}{p Y}= a_K$.

Then it follows that: 
\begin{align}
\frac{p_n X_n}{\pi_n q_n} = \frac{a_n}{\sum_{i=0}^{n-1} a_i} 
= \frac{\alpha_n}{1- \alpha_n} \left( \frac{p_n}{\pi_n} \right)^{\gamma_n}
\label{xxx}
\end{align}
The $n$th factor cost share is by definition $a_{n}=\frac{p_n X_n}{p Y}$.
By defining ${z}_n=\frac{a_n}{\sum_{i=0}^{n-1}a_i}$ and $\zeta_n = \frac{\alpha_n}{1-\alpha_n}$, taking the natural log and indexing samples by $t$, we have the following equation similar to (\ref{regtwo}):
\begin{align}
\ln {{z}}_{nt} = \ln \zeta_n + {\gamma_n} \ln p_{nt}/\pi_{nt}  + \varepsilon_{nt}
\label{regnest}
\end{align}
Equation (\ref{regnest}) is a simple regression equation, except that $\pi_{nt}$ is not an observed data but a state variable evaluated by the nest one stage inside.
In other words, parameters $(\zeta_n, \gamma_n)$ are estimated conditional on $\pi_{nt}$, which is given by the nest dual function whose parameters $(\zeta_{n-1}, \gamma_{n-1})$ are estimated conditional on $\pi_{n-1\,t}$, and so on.
The final standing state variable is available, nonetheless, i.e., $\pi_{0t} = w_{0t}$ for all $t$.

This property leads one to consider the following dynamic optimization problem, that is to
\begin{align}
\min_{\zeta_{n}, \gamma_{n}}~\sum_{n=0}^N \sum_{t=0}^T \left( \ln {{z}}_{nt} - \ln \zeta_n - {\gamma_n} \ln p_{nt}/\pi_{nt} \right)^2
\end{align}
subject to the ($n$-wise) state transition specified by the CES dual function (\ref{ncf}), in a following form:
\footnote{Here, we describe nest unit cost (CES dual) in terms of state 
$\bm{\pi}_n=(\pi_{n0}, \cdots, \pi_{nT})$ 
and controls $(\zeta_n, \gamma_n)$.
 }
\begin{align}
\pi_{n+1 t}
=
c_{n} \left( p_{nt}, \pi_{nt}; \zeta_n, \gamma_n \right)
=
\left( \frac{\zeta_{n}}{1+\zeta_{n}} (p_{n t})^{\gamma_{n}} + \frac{1}{1+\zeta_{n}} (\pi_{n t})^{\gamma_{n}} \right)^\frac{1}{\gamma_{n}}
\end{align}
We can interpret this problem by the following Bellman equation:
\begin{align}
\mathcal{V}(\bm{\pi}_{n}) = 
\min_{\zeta_{n}, \gamma_{n}} \mathcal{O}_n \left( \zeta_n, \gamma_n; \bm{\pi}_n \right) + 
\mathcal{V}\left( \mathcal{T}_n \left(\bm{\pi}_n; \zeta_{n}, \gamma_{n} \right) \right)
\label{bellman}
\end{align}
The objective function and the state transition can be specified as follows:
\begin{align}
\mathcal{O}_n\left( \zeta_n, \gamma_n; \bm{\pi}_n \right) 
&= \sum_{t=0}^T \left( \ln {{z}}_{nt} - \ln \zeta_n - {\gamma_n} \ln p_{nt}/\pi_{nt} \right)^2
\\
\mathcal{T}_n\left(\bm{\pi}_n; \zeta_{n}, \gamma_{n} \right)
&=
\left( c_{n} \left( p_{n0}, \pi_{n0}; \zeta_n, \gamma_n \right), \cdots, c_{n} \left( p_{nT}, \pi_{nT}; \zeta_n, \gamma_n \right) \right)
=\bm{\pi}_{n+1}
\end{align}

We consider solving this problem through backward induction.
At the terminal ($n=N$), the problem is to simply minimize the sum of squared residuals with no constraint, i.e., 
\begin{align}
\mathcal{V}(\bm{\pi}_{N})
=\min_{\zeta_N, \gamma_N} \mathcal{O}_N\left( \zeta_N, \gamma_N; \bm{\pi}_N \right) 
=\min_{\zeta_N, \gamma_N}  \sum_{t=0}^T \left( \ln {{z}}_{Nt} - \ln \zeta_N - {\gamma_N} \ln p_{Nt}/\pi_{Nt} \right)^2
\label{valueterm}
\end{align}
Below are the solutions (OLS estimators) of (\ref{valueterm}):
\begin{align}
\hat{\gamma}_N = \frac{\Cov(\ln {z}_{Nt}, \ln p_{Nt}/\pi_{Nt})}{\Var(\ln p_{Nt}/\pi_{Nt})}
&&
\ln \hat{\zeta}_N
= \frac{\sum_{t=0}^T\ln {{z}_{Nt}} - \hat{\gamma} \sum_{t=0}^T{\ln p_{Nt}/\pi_{Nt}}}{T+1} 
\end{align}
Using these parameters, the final value function can be reduced as follows:
\begin{align}
\mathcal{V}(\bm{\pi}_{N})
= \sum_{t=0}^T 
\left( \ln {{z}}_{Nt} - \ln \hat{\zeta}_N - {\hat{\gamma}_N} \ln p_{Nt}/\pi_{Nt} \right)^2
\end{align}
 To solve the Bellman equation further backward, this value function must be evaluated, perhaps numerically.
However, this task can become quite laborious, especially when the state variables' dimension $T$ is large.

\subsection{Restoring Parameters}
The value function becomes null, i.e., $\mathcal{V}\left( \bm{\pi}_n \right) = 0$ for $n=N, N-1, \cdots, 1, 0$, in the case of a two-point regression, i.e., $T=1$.
In this case, the parameters of (\ref{bellman}) can be solved, for all $n$ as follows:
\begin{align}
&\bar{\gamma}_n = \frac{\ln {{z}}_{n1} -\ln {{z}}_{n0}}{\ln p_{n1}/\pi_{n1} - \ln p_{n0}/\pi_{n0}} 
&\ln \bar{\zeta}_n = \frac{\ln {{z}}_{n0} \ln p_{n1}/\pi_{n1} -\ln {{z}}_{n1} \ln p_{n0}/\pi_{n0}}{\ln p_{n1}/\pi_{n1} - \ln p_{n0}/\pi_{n0}} 
\label{parmcces}
\end{align}
Below, we verify this by induction.
First, we know by the nature of a two-point regression that the terminal value function reduces to zero; that is, by substituting (\ref{parmcces}) into (\ref{valueterm}), we have the following:
\begin{align}
\mathcal{V}(\bm{\pi}_{N})
= \sum_{t=0}^1 \left( \ln {{z}}_{Nt} - \ln \bar{\zeta}_N - {\bar{\gamma}_N} \ln p_{Nt}/\pi_{Nt} \right)^2
=0
\label{vn}
\end{align}
Similarly, all error terms of a two-point regression must vanish for all $n$:
\begin{align}
\mathcal{O}_n\left( \bar{\zeta}_n, \bar{\gamma}_n; \bm{\pi}_n \right) 
= \sum_{t=0}^1 \left( \ln {{z}}_{nt} - \ln \bar{\zeta}_n - {\bar{\gamma}_n} \ln p_{nt}/\pi_{nt} \right)^2
= 0
\label{omegazero}
\end{align}
By virtue of a Bellman equation (\ref{bellman}), all value functions reduce to zero, with terminal condition (\ref{vn}), i.e.,
\begin{align}
\mathcal{V}(\bm{\pi}_{n}) =\mathcal{O}_{n}\left( \bar{\zeta}_{n}, \bar{\gamma}_{n}; \bm{\pi}_{n} \right) + \mathcal{V}(\bm{\pi}_{n+1}) = 0
\end{align}
State variables $\bm{\pi}_n = \left( {\pi}_{n0}, {\pi}_{n1} \right)$ are recursively solved by the transition function $\mathcal{T}$ under the restoring parameters $(\bar{\zeta}_n, \bar{\gamma}_n)$ obtained by (\ref{parmcces}) for $n = 0, 1, \cdots, N$, using initial condition, i.e., $\bm{\pi}_0 = \left( \pi_{00}, \pi_{01} \right) = \left( {w}_{0}, {w}_{1} \right)$, as follows: 
\begin{align}
\mathcal{T}_{n}\left(\cdots \mathcal{T}_1\left( \mathcal{T}_0\left( \bm{\pi}_0; \bar{\zeta}_0, \bar{\gamma}_0 \right); \bar{\zeta}_1, \bar{\gamma}_1 \right) \cdots \right) = \bm{\pi}_{n+1}
\end{align}

Finally, we can calibrate the changes in TFP when output price $(p_0, p_1)$ can be monitored.
Because a zero-profit condition implies $p_t Y_t = \pi_t Q_t$, where $p_t$ and $Y_t$ denote the observed price and output at $t$, respectively, we shall introduce $\theta_t$ such that 
\begin{align}
Y_0/Q_0 = \theta_0 = \pi_0/p_0
&&
Y_1/Q_1 = \theta_1 = \pi_1/p_1
\end{align}
From the first identity in both periods, we know that $\theta_t$ is TFP.
Thus, TFP growth ($\text{TFPg}=\Delta \ln \theta = \ln \theta_1/\theta_0$) can be calibrated by the growth of unit cost less the growth of the observed output price, namely,
\begin{align}
\text{TFPg} 
= \ln {\theta_1}/{\theta_0} = \ln {{\pi}_1}/{{\pi}_0} - \ln {p_1}/{p_0}
\label{tfpg}
\end{align}
Figure \ref{fig_tfpg} displays sectoral TFPg, measured by cascaded CES production under the restoring parameters specified in (\ref{parmcces}).
Factor substitution elasticities of a cascaded CES production is examined in \hyperlink{app}{Appendix}.

\subsection{Structural Transformation Restoration}
We reconfirm that parameters (\ref{parmcces}) are 
state-restoring (i.e., two observed cost shares are completely restored by the observed factor prices) as (\ref{omegazero}) and (\ref{xxx}) indicate the following for all $n$:
\begin{align}
\frac{a_{n0}}{\sum_{i=0}^{n-1} a_{i0}}= \frac{\bar{\alpha}_n}{1-\bar{\alpha}_n} \left( \frac{p_{n0}}{\pi_{n0}}\right)^{\bar{\gamma}_n}
&&
\frac{a_{n1}}{\sum_{i=0}^{n-1} a_{i1}}= \frac{\bar{\alpha}_n}{1-\bar{\alpha}_n} \left( \frac{p_{n1}}{\pi_{n1}}\right)^{\bar{\gamma}_n}
\label{ant}
\end{align}
Notice that $\pi_{nt}$ can be obtained as a function of $\left( w_{t}, r_{t}, p_{1t}, \cdots, p_{n-1 \, t} \right)$ with regard to (\ref{ncf}) given the parameters ($\bar{\alpha}_0, \cdots, \bar{\alpha}_{n-1}$, $\bar{\gamma}_0, \cdots, \bar{\gamma}_{n-1}$).
With regard to (\ref{ant}), the observed cost shares ($a_{Lt}, a_{Kt}, a_{1t}, \cdots, a_{Nt}$) for the two periods ($t=0,1$), where $a_{Kt}$ corresponds to $a_{0t}$ and $a_{Lt} = 1-a_{Kt}-a_{1t}- \cdots -a_{Nt}$, will be restored by the corresponding factor prices ($w_{t}$, $r_{t}$, $p_{1t}$, $\cdots$, $p_{Nt}$) under the parameters ($\bar{\alpha}_0, \cdots, \bar{\alpha}_{N}$, $\bar{\gamma}_0, \cdots, \bar{\gamma}_{N}$) that are given by the observed cost shares and factor prices for the two periods, as given in (\ref{parmcces}). 

Below we investigate this property from another perspective.
As regards the Shephard's lemma, cost shares can always be written in terms of the unit cost function, as follows:
\begin{align}
\frac{\partial C_j \left( \bm{p}, r, w \right)}{\partial p_i} \frac{p_i}{\theta_j p_j} = a_{ij} 
&&
\frac{\partial C_j \left( \bm{p}, r, w  \right)}{\partial r} \frac{r}{\theta_j p_j} = a_{Kj}
&&
\frac{\partial C_j \left( \bm{p}, r, w  \right)}{\partial w} \frac{w}{\theta_j p_j} = a_{Lj}
\label{aij}
\end{align}
where, we now index sectors explicitly by $j = 1, \cdots, N$ and intermediate factor inputs by $i =1, \cdots, N$. 
We write down the above more concisely as follows:
\begin{align}
\left< \bm{p} \right> \nabla \bm{C}
\left< \bm{\theta} \right>^{-1}\left< \bm{p} \right>^{-1} = \bm{A}
&&
r \bm{C}_r 
\left< \bm{\theta} \right>^{-1}\left< \bm{p} \right>^{-1} = \bm{a}_{K}
&&
w \bm{C}_w
\left< \bm{\theta} \right>^{-1}\left< \bm{p} \right>^{-1} = \bm{a}_{L}
\end{align}
Angle brackets indicate diagonalization. 
$\bm{A}$ can otherwise be recognized as the input coefficient matrix.
Furtermore, $\bm{a}_L$ may be referred to as labor intensity.
We know from (\ref{ant}) that under the restoring parameters ($\bar{\alpha}_{j0}$, $\cdots$, $\bar{\alpha}_{jN}$, $\bar{\gamma}_{j0}$, $\cdots$, $\bar{\gamma}_{jN}$), the system of unit cost functions $\bm{C}$ traces the observed cost-share structures of the two periods ($\bm{A}_0$, $\bm{a}_{K0}$, $\bm{a}_{L0}$, $\bm{A}_1$, $\bm{a}_{K1}$, $\bm{a}_{L1}$) in terms of prices of all commodities $\left( \bm{p}_0, r_0, w_0, \bm{p}_1, r_1, w_1 \right)$ and TFPs $\left( \bm{\theta}_0, \bm{\theta}_1 \right)$.
We write this explicitly below:
\begin{align}
\left< \bm{p}_0 \right> \nabla \bm{C}
\left< \bm{\theta}_0 \right>^{-1}\left< \bm{p}_0 \right>^{-1} = \bm{A}_0
&&
r_0 \bm{C}_r 
\left< \bm{\theta}_0 \right>^{-1}\left< \bm{p}_0 \right>^{-1} = \bm{a}_{K0}
&&
w_0 \bm{C}_w
\left< \bm{\theta}_0 \right>^{-1}\left< \bm{p}_0 \right>^{-1} = \bm{a}_{L0} \label{eq0}
\\
\left< \bm{p}_1 \right> \nabla \bm{C}
\left< \bm{\theta}_1 \right>^{-1}\left< \bm{p}_1 \right>^{-1} = \bm{A}_1
&&
r_1 \bm{C}_r 
\left< \bm{\theta}_1 \right>^{-1}\left< \bm{p}_1 \right>^{-1} = \bm{a}_{K1} 
&&
w_1 \bm{C}_w
\left< \bm{\theta}_1 \right>^{-1}\left< \bm{p}_1 \right>^{-1} = \bm{a}_{L1} 
\label{eq1}
\end{align}

The equilibrium price of commodities $\bm{p}$ is the fixed point of the following feedback system:
\begin{align}
\bm{p} = \bm{C} \left( \bm{p}, r, w \right) \left< \bm{\theta} \right>^{-1}
\label{feedback}
\end{align}
We solve for the fixed point of (\ref{feedback}) by recursively plugging $\bm{p}$ from the left-hand side into that from the right-hand side.
This procedure will globally converge into a unique fixed point (i.e., equilibrium price) as long as we can be assured that (\ref{feedback}) is a contraction mapping \citep{kras} with respect to $\bm{p}$. 
We know, however, that (\ref{feedback}) is a contraction mapping since $C_j ({p_1, \cdots, p_N}, r, w)$ is monotonic and homogeneous of degree one with respect to $({p_1, \cdots, p_N}, r, w)$, for all $j$.
In that case, for any $0<({p_1, \cdots, p_N}, r, w)$, we can always find a large $K > 1$ and a small $0< k<1$ such that $k p_j < C_j ({k p_1, \cdots, k p_N}, r, w)/\theta_j < C_j ({K p_1, \cdots, K p_N}, r, w)/\theta_j < K p_j$, for all $j$, so that the rage $( k {p}_j, K {p}_j)$ for all $j$ is contracting by the mapping (\ref{feedback}).

Under the restoring parameters, $\left( \bm{p}_0, \bm{p}_1 \right)$ must be the fixed point solutions given the price of primary factors $\left( r_0, w_0, r_1, w_1\right)$ and TFPs $\left( \bm{\theta}_0, \bm{\theta}_1 \right)$.
We write this explicitly below:
\begin{align}
\bm{p}_0 = \bm{C} \left( \bm{p}_0, r_0, w_0 \right) \left< \bm{\theta}_0 \right>^{-1}
&&
\bm{p}_1 = \bm{C} \left( \bm{p}_1, r_1, w_1 \right) \left< \bm{\theta}_1 \right>^{-1}
\label{fp}
\end{align}
Hence, regarding (\ref{eq0}--\ref{eq1}) and (\ref{fp}), we know that primary factor prices $\left( r_0, w_0, r_1, w_1\right)$ and TFPs $\left( \bm{\theta}_0, \bm{\theta}_1 \right)$ lead to the observed cost-share structures (input coefficient matrices) of the two periods, namely, ($\bm{A}_0$, $\bm{a}_{K0}$, $\bm{a}_{L0}$, $\bm{A}_1$, $\bm{a}_{K1}$, $\bm{a}_{L1}$), under the restoring parameters.

\subsection{Total Factor Productivity}

\begin{figure}[t!]
\centering
  \begin{minipage}[b]{0.475\textwidth}
    \includegraphics[width=\textwidth]{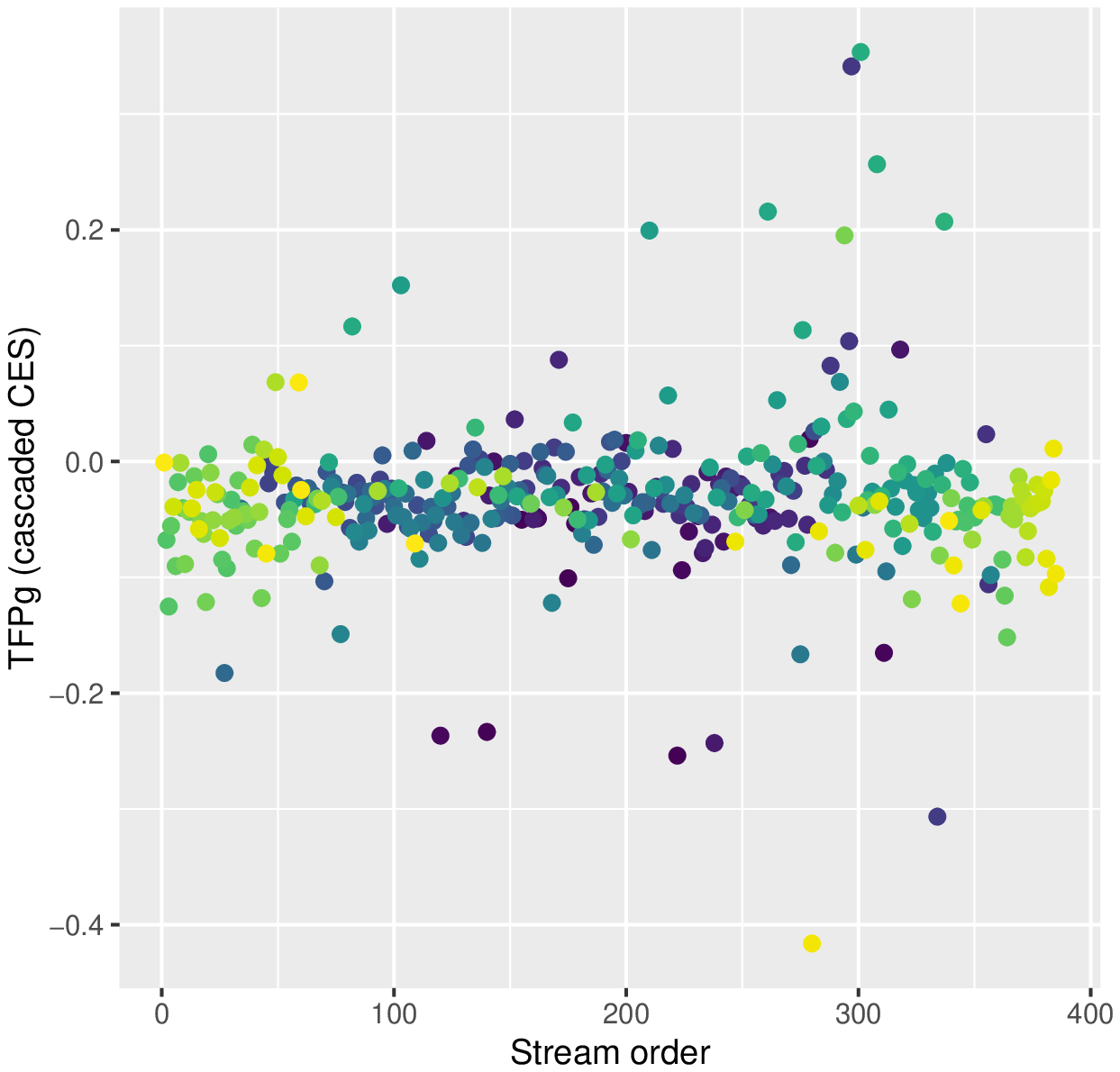}
    \caption{Sectoral TFPg calibration based on cascaded CES function with restoring parameters. 
    Colors correspond to the classification order of sectors.}
   \label{fig_tfpg}
  \end{minipage}
\hspace{0.03\textwidth}
   \begin{minipage}[b]{0.475\textwidth}
    \includegraphics[width=\textwidth]{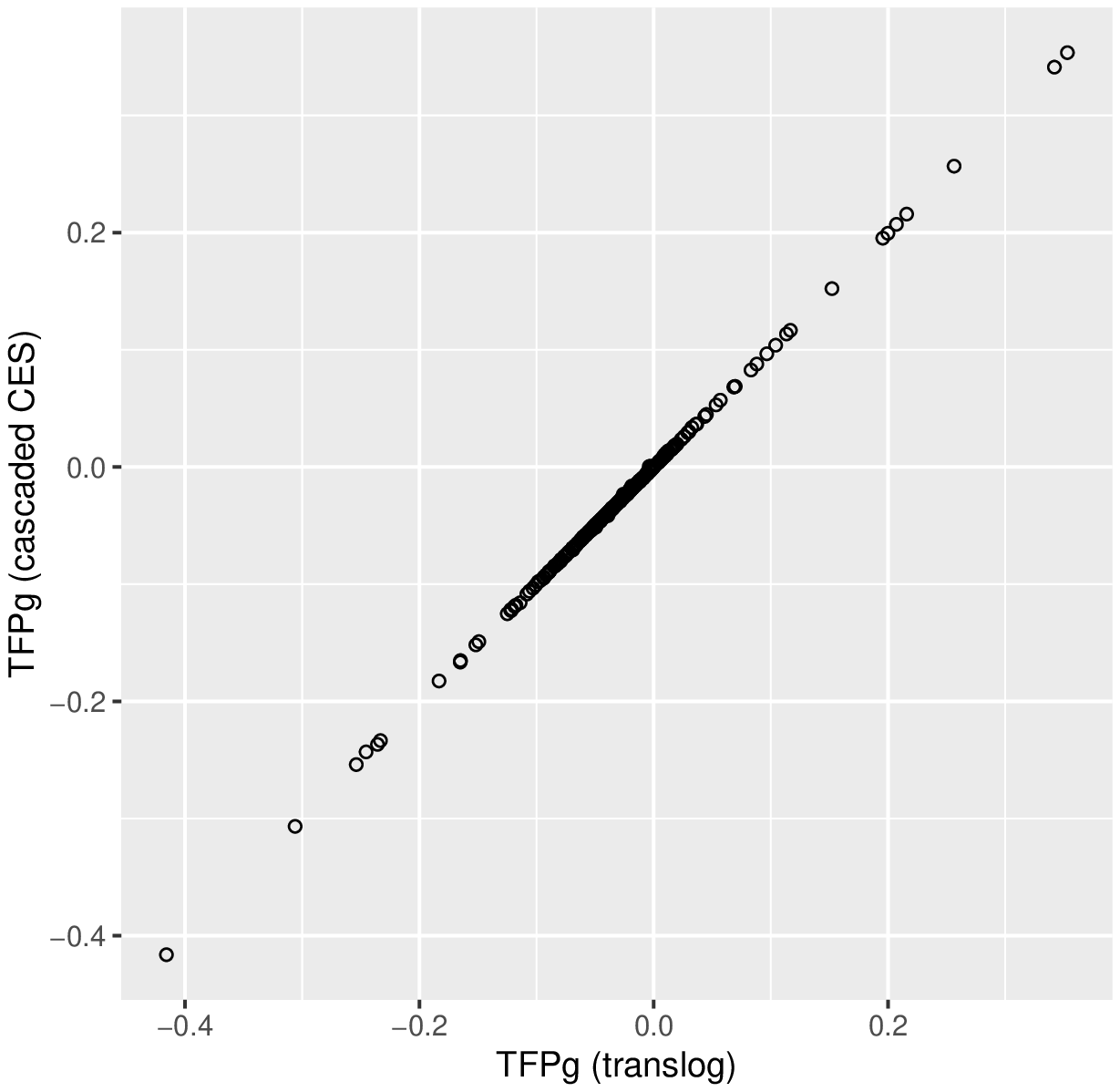}
    \caption{Correspondences between TFPg of cascaded CES functions with restoring parameters and the logarithm of T{\"o}rnqvist Index (labeled TFPg (translog)). }
   \label{fig_tfpgcon}
  \end{minipage}
\end{figure}

As mentioned earlier, sector-wise TFPg of cascaded CES production evaluated using (\ref{tfpg}) under the restoring parameters is shown in Figure \ref{fig_tfpg}.
In Figure \ref{fig_tfpgcon}, we display the same result in reference to the log of T{\"o}rnqvist Index for each sector, obtainable by the following formula (where index $j$ is omitted):
\begin{align}
\text{TFPg (translog)} = 
  \sum_{i=1}^N \frac{a_{i0} +a_{i1}}{2} \ln \frac{p_{i1}}{p_{i0}}
+ \frac{a_{L0} + a_{L1}}{2} \ln \frac{w_1}{w_0}
+ \frac{a_{K0} + a_{K1}}{2} \ln \frac{r_1}{r_0}
- \ln \frac{p_{1}}{p_{0}} 
\end{align}
\citet{diewert} showed that this log of T{\"o}rnqvist Index exactly captures the underlying translog function's productivity growth, which potentially materializes the two observed cost shares, even without knowing the parameters.
Notice that TFPg (cascaded CES) and TFPg (translog) show extreme concordances.

\subsection{Structural Transformation Interpolation}
\begin{figure}[t!]
\centering
  \begin{minipage}[b]{0.475\textwidth}
    \includegraphics[width=\textwidth]{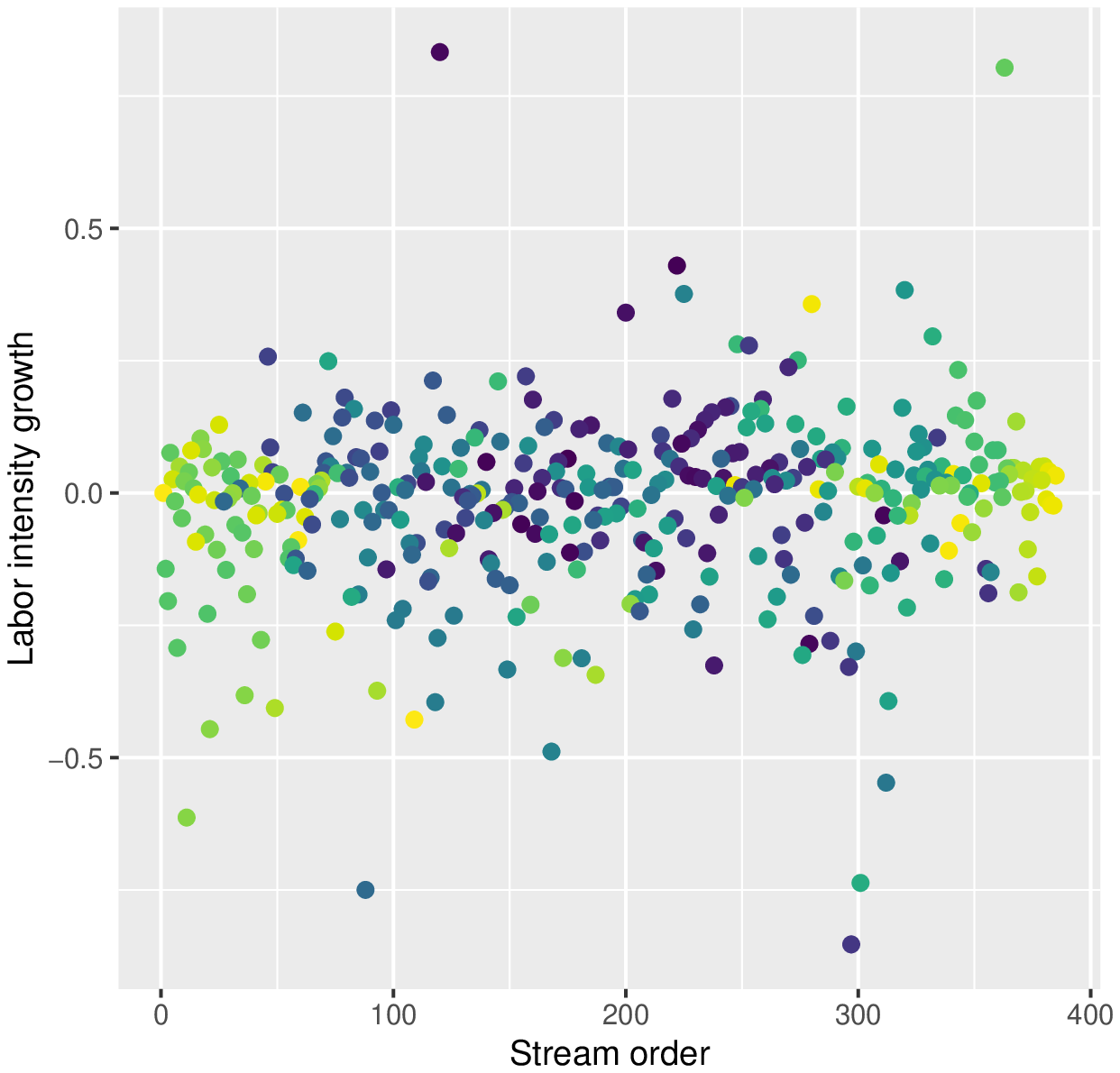}
    \caption{Sectoral labor intensity growth as recorded in linked input--output tables. 
Same results are obtainable via cascaded CES functions with restoring parameters.
}
   \label{fig_tfpg0.5}
  \end{minipage}
\hspace{0.03\textwidth}
   \begin{minipage}[b]{0.475\textwidth}
    \includegraphics[width=\textwidth]{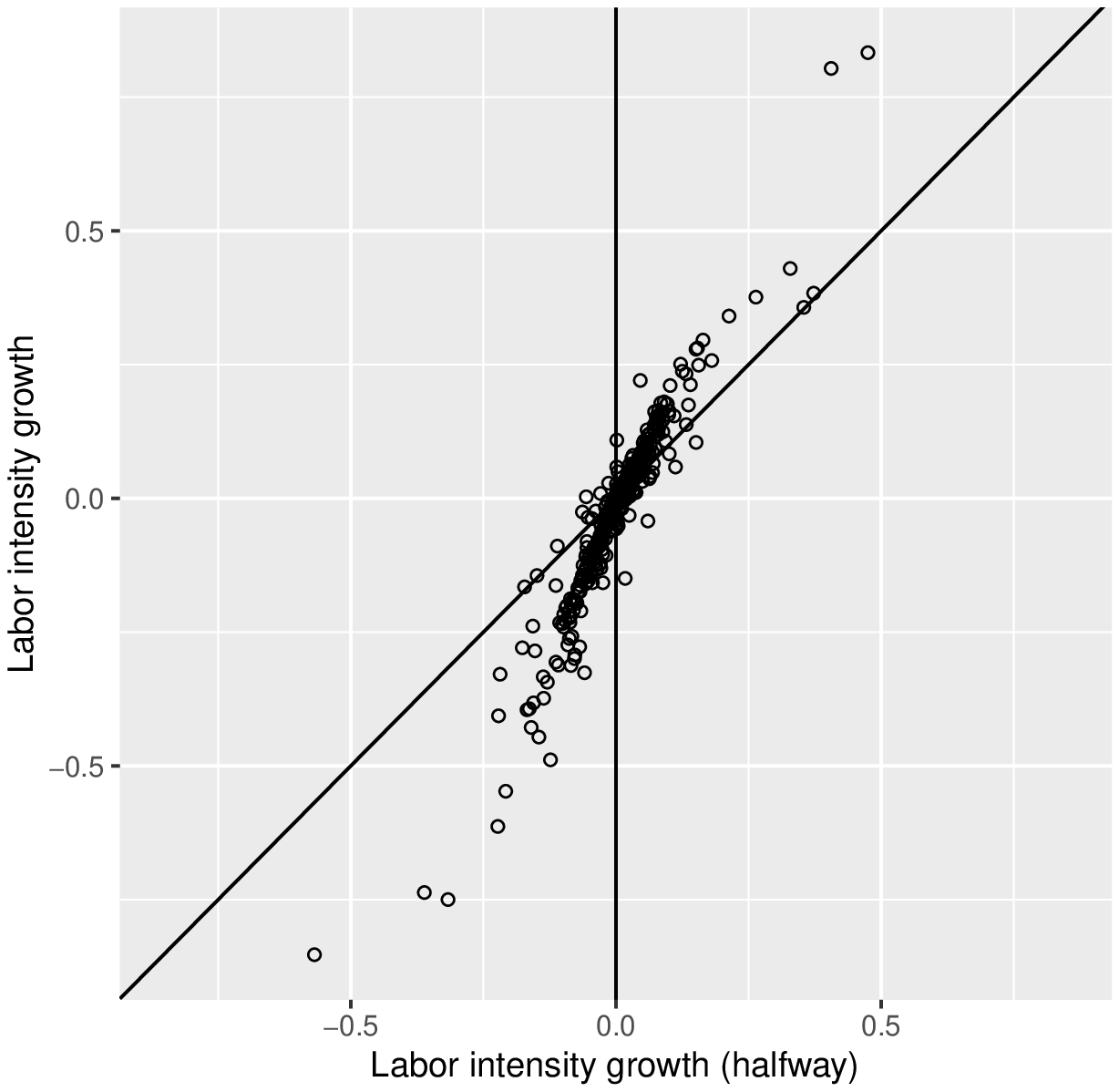}
    \caption{Correspondences between labor intensity growth and halfway labor intensity growth. The 45-degree line from the vertical line indicates full labor intensity growth.}
    \label{fig_tfpg0.5dy}
  \end{minipage}
\end{figure}

Let us consider the following labor intensity growth between the two periods $t=0,1$ for the $j$th sector:
\begin{align}
\text{Labor intensity growth}
= \Delta \ln a_{Lj}
= \ln \frac{a_{Lj1}}{a_{Lj0}}
\end{align}
Our database allows us to calculate labor intensity $a_{Ljt} = ({w_{t} L_{jt}})/({p_{jt} Y_{jt})}$ for all $j$ sectors.
However, they are also replicable by the system of unit cost functions $\bm{C}$ with restoring parameters as indicated in (\ref{eq0}--\ref{eq1}).
We may then consider the following halfway labor intensity growth:
\begin{align}
\text{Halfway labor intensity growth}
= \ln \frac{a_{Lj 0.5}}{a_{Lj0}}
\end{align}
where ${a}_{Lj0.5}$ is obtained by the following calculations:
\begin{align}
\bm{a}_{L0.5} 
=
w_{0.5} \bm{C}_w
\left< \bm{\theta}_{0.5} \right>^{-1}\left< \bm{p}_{0.5} \right>^{-1} 
&&
\bm{p}_{0.5} = \bm{C} \left( \bm{p}_{0.5}, r_{0.5}, w_{0.5} \right) \left< \bm{\theta}_{0.5} \right>^{-1}
\end{align}
Here, sectoral TFP and primary factor prices are at halfway, i.e., $r_{0.5} = \frac{r_1 + r_0}{2}$, $w_{0.5} = \frac{w_1 + w_0}{2}$, and $\theta_{j0.5} = \frac{\theta_{j1} + \theta_{j0}}{2}$.
Halfway commodity equilibrium price $\bm{p}_{0.5}$ is obtained by feedback recursion.
Figure \ref{fig_tfpg0.5} displays sectoral labor intensity growth.
Figure \ref{fig_tfpg0.5dy} displays halfway labor intensity growth on the horizontal axis. 

\section{Representative Household}
\subsection{Indirect Utility Function}
We now consider an immortal representative household with identical $N$-commodity CES utility function $U(H_1,\cdots,H_N)$ of constant returns to scale, as follows:
\begin{align}
\sum_{t=0}^\infty \beta^t U\left( \bm{H}_t \right)
= \sum_{t=0}^\infty \beta^t \left( 
\sum_{i=1}^N (\mu_i)^{\frac{1}{1-\lambda}}(H_{it})^{\frac{\lambda}{\lambda -1}}
\right)^{\frac{\lambda -1}{\lambda}}
\label{obj}
\end{align}
where $0 < \beta <1$ denotes the discount (time preference) factor.
Elasticity of substitution, denoted as $1-\lambda$, is common across all goods. 
The share parameter $\mu_i$ must add up to one.
Here, we assume that total labor supply $\mathcal{L}_t = \sum_{j=1}^{N}L_{jt}$ is determined by the demand side for all $t$, and not included in the utility function.
Below is the representative household's budget constraints:
\begin{align}
\mathcal{B}_t + {s}_{t} (\rho \mathcal{K}_{t+1} - (1-\delta) \rho \mathcal{K}_{t} ) + \mathcal{R}_t = r_t \mathcal{K}_t + {w}_t \mathcal{L}_t  
\label{constr}
\end{align}
We denote the ratio between capital stock and capital service by $\rho$ so that $\rho \mathcal{K}$ designates capital stock total when $\mathcal{K}$ denotes capital service total.
Accordingly, $s$ denotes the price of capital.
The depreciation rate is denoted by ${\delta}$.
We use $\mathcal{B}$ to denote consumption expenditure total and $\mathcal{R}$ to denote the \text{net} export total.
Regarding the linked input--output tables (\ref{liot}), (\ref{constr}) describes an equivalence between total expenditures and income of a representative household with the following: $\mathcal{B}_t =\sum_{i=1}^{N} p_{it} H_{it}$,  
$\mathcal{R}_t = \sum_{i=1}^N p_{it} E_{it}$,
$\mathcal{L}_t = \sum_{j=1}^N L_{jt}$,  
$\mathcal{K}_t = \sum_{j=1}^N K_{jt}$, 
and
$s_t (\rho \mathcal{K}_{t+1} - (1-\delta) \rho \mathcal{K}_t ) = \sum_{i=1}^N p_{it} G_{it}$.

Maximization of (\ref{obj}) subject to (\ref{constr}) yields the following Euler equation:
\begin{align}
\frac{\partial U}{p_{it} \partial H_{it}} 
= \beta \frac{{s}_{t+1}\rho(1-\delta) + r_{t+1}}{{s}_t \rho} \frac{\partial U}{p_{it+1} \partial H_{it+1}}
\label{euler}
\end{align}
The CES indirect utility function must be of the following form: 
\begin{align}
V\left( p_{1t}, p_{2t}, \cdots, p_{Nt}; \mathcal{B}_t \right) 
= \frac{\mathcal{B}_t}{\left( \sum_{i=1}^N \mu_i (p_{it})^{\lambda} \right)^{\frac{1}{\lambda}}} 
= \frac{\mathcal{B}_t}{I\left( \bm{p}_t \right)}
\end{align} 
Here, price index $I$ as defined above, is the cost of living index.
We then know that the marginal utility of money can be expanded as follows:
\begin{align}
\frac{\partial V}{\partial \mathcal{B}_t} = \frac{\partial U}{p_{it}\partial H_{it}} = \frac{1}{I \left(\bm{p}_t \right)}
\label{roy1}
\end{align}
Hence, by (\ref{euler}) and (\ref{roy1}), the following condition must hold.
\begin{align}
\beta \frac{{s}_{t+1} \rho (1-\delta) + r_{t+1}}{ {s}_t \rho} = \frac{ I(\bm{p}_{t+1})}{ I(\bm{p}_{t})}
\label{dy}
\end{align}

\subsection{Estimation of Parameters}
For a CES utility, Roy's identity indicates the following:
\begin{align}
\frac{\partial V}{\partial p_i}/\frac{\partial V}{\partial \mathcal{B}} = - H_i 
= - \frac{\mathcal{B}}{\sum_{j=1}^N \mu_j (p_j)^{\lambda}} \mu_i (p_i)^{\lambda-1} 
\label{roy}
\end{align}
The expenditure share of item $i$ can be found by rearranging the above terms as follows:
\begin{align}
b_i
= \frac{p_i H_i}{\mathcal{B}} 
= \frac{\mu_i (p_i)^{\lambda}}{\sum_{j=1}^N \mu_j (p_j)^{\lambda}} 
= {\mu_i \left( \frac{p_i}{I(\bm{p})} \right)^\lambda }
\label{bmu}
\end{align}
Observing item-wise expenditure shares $b_{i}$ in our data as above, (\ref{bmu}) can be expanded with an error term to obtain the following regression equation:
\begin{align}
\ln b_{it} = \ln \mu_{i} - \lambda \ln I_{t} + \lambda \ln p_{it} + \varepsilon_{it}
\end{align}
where we index observed samples by $t$.

We thus find that $\lambda$ can be estimated via fixed effect or by the following simple regression:
\begin{align}
\Delta \ln b_i =  - \lambda \Delta \ln I + \lambda \Delta \ln p_i + \Delta \varepsilon_i
\label{fe}
\end{align}
Here, $\Delta$ indicates differences between observations for $t=0,1$.
Upon estimation of (\ref{fe}), at least two issues must be dealt with.
One is the endogeneity of the regressor, and the other is the heteroskedasticity of the error term.
Regarding endogeneity (i.e., endogeneity by the anticipated reverse causality that a representative household's expenditures can affect commodity prices), 
we perform instrumental variables estimation using sector-wise (thus, commodity-wise) cascaded CES productivity growths $\Delta \ln \theta_i$ measured previously as instruments which must strongly correlate with the regressor but not with the error term.

Regarding heteroskedasticity, we consider potential measurement errors for log-difference transformation of two stochastic variables ($b_{i0}$, $b_{i1}$). 
We suppose that these are normally distributed random variables with mean $(b_{i0}, b_{i1})$ and a homoskedastic variance $\sigma^2$.
In this case the dependent variable's variance can be approximated as described below:
\begin{align}
\Var \left[ \Delta \ln b_{i} \right]
=
\Var \left[ \ln b_{i1} - \ln b_{i0} \right] \approx 
 \sigma^2 \left( \frac{1}{(b_{i1})^2} + \frac{1}{(b_{i0})^2} \right) 
=\sigma^2 (m_i)^2
\end{align} 

We therefore perform a weighted two-stage least squares estimation using $m_i$ (defined above) as weights and $\Delta \ln \theta_i$ and $\exp \left( \Delta \ln \theta_i \right)$ as instruments.
The result is shown below (with standard errors in parentheses):
\begin{align}
\Delta \ln b_i = \underset{(0.00850)}{0.00561} + \underset{(0.35218)}{1.09631} \Delta \ln p_i 
\label{result}
\end{align}
Considering the first-stage F statistic ($\text{F}(2, 265)= 119.57$) we are not concerned about a weak instrument. 
Regarding the Durbin--Wu--Hausman test for regressor endogeneity (Durbin $\chi^2 (1) = 10.5032$, Wu-Hausman F(1, 265) $=10.8093$), we reject the null hypothesis that the regressor is exogenous.
In testing the instruments for overidentifying restrictions (Sargan $\chi^2(1) =0.2917$, Basmann $\chi^2(1) = 0.2887$), we do not reject the null hypothesis that at least one of the instruments is endogenous.

For empirical study, we standardize prices at $t=1$ so that $p_{11}=\cdots=p_{N1}=w_1 =r_1 =1$.\footnote{As mentioned earlier, this is equivalent to normalizing the empirical CES function at $t=1$.}
In this event, the price index also is standardized as $I=1$ regardless of the value of $\lambda$, since $\sum_{i=1}^N \mu_i =1$.
Then, according to (\ref{bmu}), we know that $\mu_i = b_{i1}$ for all $i$.
With the estimated substitution elasticity (${\lambda}=1.09631$), the price index function (normalized at $t=1$) becomes as follows:
\begin{align}
I\left( \bm{p} \right)
=\left( 
b_{11} (p_{1})^{{\lambda}} + \cdots + b_{N1} (p_{N})^{{\lambda}}
\right)^{\frac{1}{{\lambda}}}
\end{align}

\section{Application}
\subsection{Structural Transformation Extrapolation }
Below are the representative household's budget constraints at $t = 0, 1$:
\begin{align}
\mathcal{B}_0 + {s}_{0} (\rho \mathcal{K}_{1} - (1-\delta) \rho \mathcal{K}_{0} ) + \mathcal{R}_0 &= r_0 \mathcal{K}_0 + {w}_0 \mathcal{L}_0 \label{bu0}
\\
\mathcal{B}_1 + {s}_{1} (\rho {K}_{2} - (1-\delta) \rho \mathcal{K}_{1} ) + \mathcal{R}_1 &= r_1 \mathcal{K}_1+ {w}_1 \mathcal{L}_1   \label{bu1}
\end{align}
In reference to the linked input--output tables (\ref{liot}), the right-hand sides represent total value added, and the left-hand sides represent total final demand.
Here, the prices of capital service formation i.e., $s_0 \rho$ and $s_1 \rho$ are not observed directly.
We may however estimate $s_0 \rho$ as we employ $\delta$ from external sources\footnote{We use a five-year capital depreciation rate $\delta = 1 - (1 - 0.125)^5$ in reference to \cite{nomurasuga}.} and employ $K_0$ and $K_1$ of the left-hand side of (\ref{bu0}) from the right-hand sides of (\ref{bu0}) and (\ref{bu1}).
We display below the dynamic first-order condition (\ref{dy}) for the two periods:
\begin{align}
\beta \frac{{s}_{1} \rho (1-\delta) + r_{1}}{{s}_0 \rho} = \frac{ I(\bm{p}_{1})}{ I(\bm{p}_{0})} \label{euler01}
\end{align}
Since we employ $\beta$ from external sources, we can use (\ref{euler01}) to evaluate $s_1 \rho$.
\footnote{We use a five-year time preference factor $\beta = (1 + 0.03)^{-5}$ in reference to \cite{kawasaki, idagoto}.}

Let an alternative equilibrium at $t=1$, with alternative productivity $\bm{\theta}_1^\prime$, hereafter be indicated by a \textit{prime}.
The alternative equilibrium price $\bm{p}_1^\prime$ is specified as the fixed point of the following feedback system:
\begin{align}
\bm{p}_1^\prime = \bm{C} \left( \bm{p}_1^\prime, r_1, w_1 \right) \left< \bm{\theta}_1^\prime \right>^{-1}
\label{xeq}
\end{align}
Assume that primary factor prices $(r_1, w_1)$ without feedbacks are fixed in the alternative equilibrium.
Furthermore, assume for simplicity that capital service $\mathcal{K}_1$ and the net export total $\mathcal{R}_1^\prime = \sum_{i=1}^N p_{i1}^\prime E_{i1}^\prime=\sum_{i=1}^N p_{i1} E_{i1} = \mathcal{R}_1$ are also fixed.\footnote{In other words, we assume that price elasticity of exports (and imports) are all one for all $i$.}
The household's budget constraint then becomes as follows:
\begin{align}
\mathcal{B}_1^\prime + {s}_{1}^\prime (\rho \mathcal{K}_{2}^\prime - (1-\delta) \rho \mathcal{K}_{1} ) + \mathcal{R}_1 &= r_1 \mathcal{K}_1 + {w}_1 \mathcal{L}_1^\prime
\label{postbal}
\end{align}
In order to estimate $s_1^\prime \rho$ we apply (\ref{euler01}) with the following modification:
\begin{align}
\frac{{s}_{1}^\prime \rho (1-\delta) + r_{1}}{{s}_{1} \rho (1-\delta) + r_{1}} = \frac{ I(\bm{p}_{1}^\prime)}{ I(\bm{p}_{1})}
\label{xx}
\end{align}
We further evaluate the alternative fixed capital formation (the second term of (\ref{postbal})) using the price elasticity of fixed capital formation $\eta$, which we measure below.
\begin{align}
\eta
&=
\frac{(\mathcal{K}_2 - (1 - \delta) \mathcal{K}_1) - (\mathcal{K}_1 - (1 - \delta) \mathcal{K}_0)}{s_1 \rho - s_0 \rho} \frac{s_0 \rho}{\mathcal{K}_1 - (1-\delta) \mathcal{K}_0}   =-0.80
\\
&=
\frac{(\mathcal{K}_2^\prime - (1 - \delta) \mathcal{K}_1) - (\mathcal{K}_1 - (1 - \delta) \mathcal{K}_0)}{s_1^\prime \rho - s_0 \rho} \frac{s_0 \rho}{\mathcal{K}_1 - (1-\delta) \mathcal{K}_0}   \label{els}
\end{align}
Once we know $\mathcal{B}_1^\prime$ from (\ref{postbal}), household consumption can be evaluated, regarding (\ref{bmu}), as follows.
\begin{align}
p_{i1}^\prime H_{i1}^\prime 
=b_i \left( {p_{i1}^\prime}/{I(\bm{p}_{1}^\prime)} \right)^{\lambda} \mathcal{B}_1^\prime
\label{alth}
\end{align} 
As regards total labor supply, we consider the demand side of labor, as indicated earlier.
Total demanded labor for the alternative equilibrium 
can be evaluated by the following Leontief inverse calculation:
\begin{align}
\mathcal{L}_1^\prime = \bm{a}_{L1}^\prime \left[ \bm{I} - \bm{A}_{1}^\prime \right]^{-1} \left< \bm{p}_1^\prime \right> \left[ 
\bm{H}_1^\prime + \bm{G}_1^\prime + \bm{E}_1^\prime
\label{altL}
\right]
\end{align}
where the extrapolated (transformed) structure can be specified as follows:
\begin{align}
\bm{A}_{1}^\prime 
= \left< \bm{p}_1^\prime \right> \nabla \bm{C} \left< \bm{\theta}_1^\prime \right>^{-1}\left< \bm{p}_1^\prime \right>^{-1} 
&&
\bm{a}_{L1}^\prime = w_1 \bm{C}_w \left< \bm{\theta}_1^\prime \right>^{-1}\left< \bm{p}_1^\prime \right>^{-1} 
\label{postL}
\end{align}
In (\ref{altL}) the alternative vector of final demand is decomposed into three parts: household consumption $\bm{H}_1^\prime$, fixed capital formation $\bm{G}_1^\prime$, and net export $\bm{E}_1^\prime$.
Here, net exports (either in total or in a vector) are assumed to be fixed, 
and fixed capital formation total is disaggregated into commodity-wise vectors by constant ratios, i.e., $g_i = {p_{i1}G_{i1}}/{\sum_{i=1}^N p_{i1} G_{i1}}$. 
Below we describe them explicitly:
\begin{align}
p_{i1}^\prime E_{i1}^\prime =p_{i1} E_{i1}
&&
p_{i1}^\prime G_{i1}^\prime = 
g_i \left[ {s}_{1}^\prime (\rho \mathcal{K}_{2}^\prime - (1-\delta) \rho \mathcal{K}_{1} ) \right]
\end{align}
The alternative equilibrium ($\mathcal{B}_1^\prime, \mathcal{L}_1^\prime$) is solved iteratively by (\ref{postbal}, \ref{alth}, \ref{altL}), using ($\mathcal{B}_1, \mathcal{L}_1$) as initial value.
Finally, we evaluate social costs and benefits of the alternative equilibrium by the followings:
\begin{align}
\text{Benefit} = V_1^\prime - V_1 
= \frac{\mathcal{B}_1^\prime}{I\left(\bm{p}_1^\prime\right)} - \frac{\mathcal{B}_1}{I\left(\bm{p}_1\right)}
&&
\text{Cost} = w_1 \mathcal{L}_1^\prime - w_1 \mathcal{L}_1
\label{wg}
\end{align}

\subsection{Productivity Shock Injection}
We virtually inject 1 billion yen of productivity shock into each industrial sector and see how much is gained in light of the extrapolated structural transformation. 
The alternative productivity at $t=1$ inclusive of that injection into sector $k$, which we denote by $\bm{\theta}_{1}^\prime (k)$, can be specified as follows:
\begin{align}
\bm{\theta}_{1}^{\prime} (k)
= \left( \theta_{11}, \cdots, \theta_{k1}^\prime, \cdots, \theta_{N1} \right)
&&
\theta_{k1}^\prime = \theta_{k1} \left( \frac{p_{k1} Y_{k1} + \text{1 billion yen}}{p_{k1} Y_{k1}} \right)
\end{align}
where $p_{k}Y_{k}$ is the monetary output of sector $k$.
We estimate the following instantaneous welfare gain of this injection according to (\ref{wg}), where we evaluate social benefit by the ex ante price index. 
\begin{align}
\text{Welfare gain}(k)
= \left( \frac{\mathcal{B}_{1}^\prime (k)}{I\left(\bm{p}_{1}^\prime (k)  \right)} - \frac{\mathcal{B}_{1}}{I\left(\bm{p}_1\right)} \right) I\left( \bm{p}_1 \right)
- \left( w_1 \mathcal{L}_1^\prime (k) - w_1 \mathcal{L}_1 \right)
\end{align}
where $\bm{p}_1^\prime(k)$ is the fixed point of the following feedback system:
\begin{align}
\bm{p}_{1}^\prime (k) = \bm{C} \left( \bm{p}_{1}^\prime (k), r_1, w_1 \right) \left< \bm{\theta}_{1}^\prime (k) \right>^{-1}
\end{align}
Note that $\mathcal{B}_1^\prime (k)$ and $\mathcal{L}_1^\prime (k)$ are the alternative budget and labor evaluated under this injection $\bm{\theta}_1^\prime(k)$.
For a welfare assessment of $\bm{\theta}_1^\prime(k)$, we use the following figure (i.e., effectiveness), which evaluates welfare gain of $k$ in terms of a cumulative discounted present value using the rate of time preference.
\begin{align}
\text{Effectiveness}(k) 
= \frac{1}{\text{1 billion yen}}\frac{\text{Welfare gain}(k)}{1 - \beta} 
\end{align}

\begin{figure}[t!]
\centering
   \begin{minipage}[b]{0.475\textwidth}
    \includegraphics[width=\textwidth]{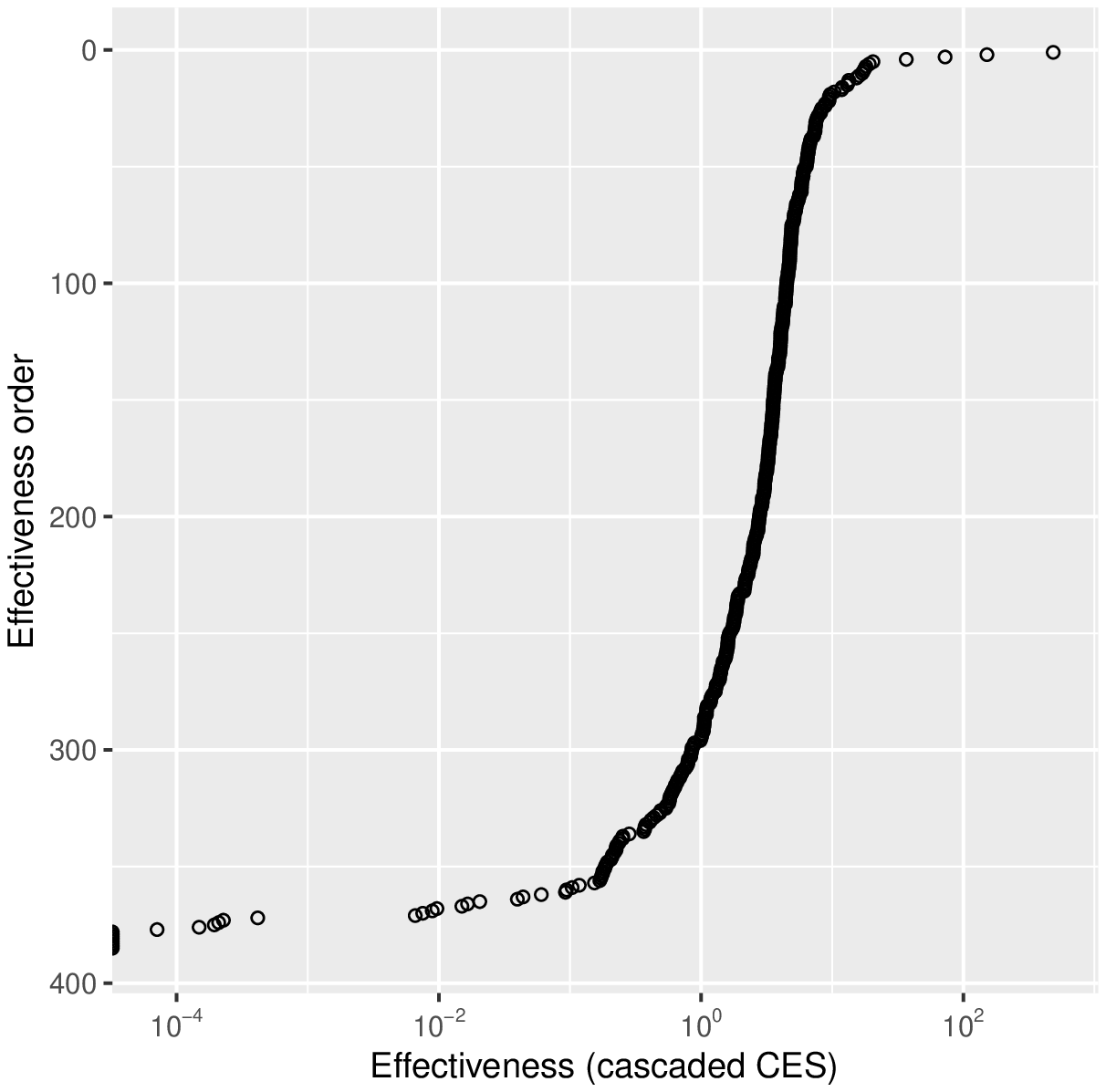}
    \caption{Sector-wise effectiveness of positive marginal productivity shock injection in a descending order.}
    \label{nine}
  \end{minipage}
\hspace{0.03\textwidth}
  \begin{minipage}[b]{0.475\textwidth}
    \includegraphics[width=\textwidth]{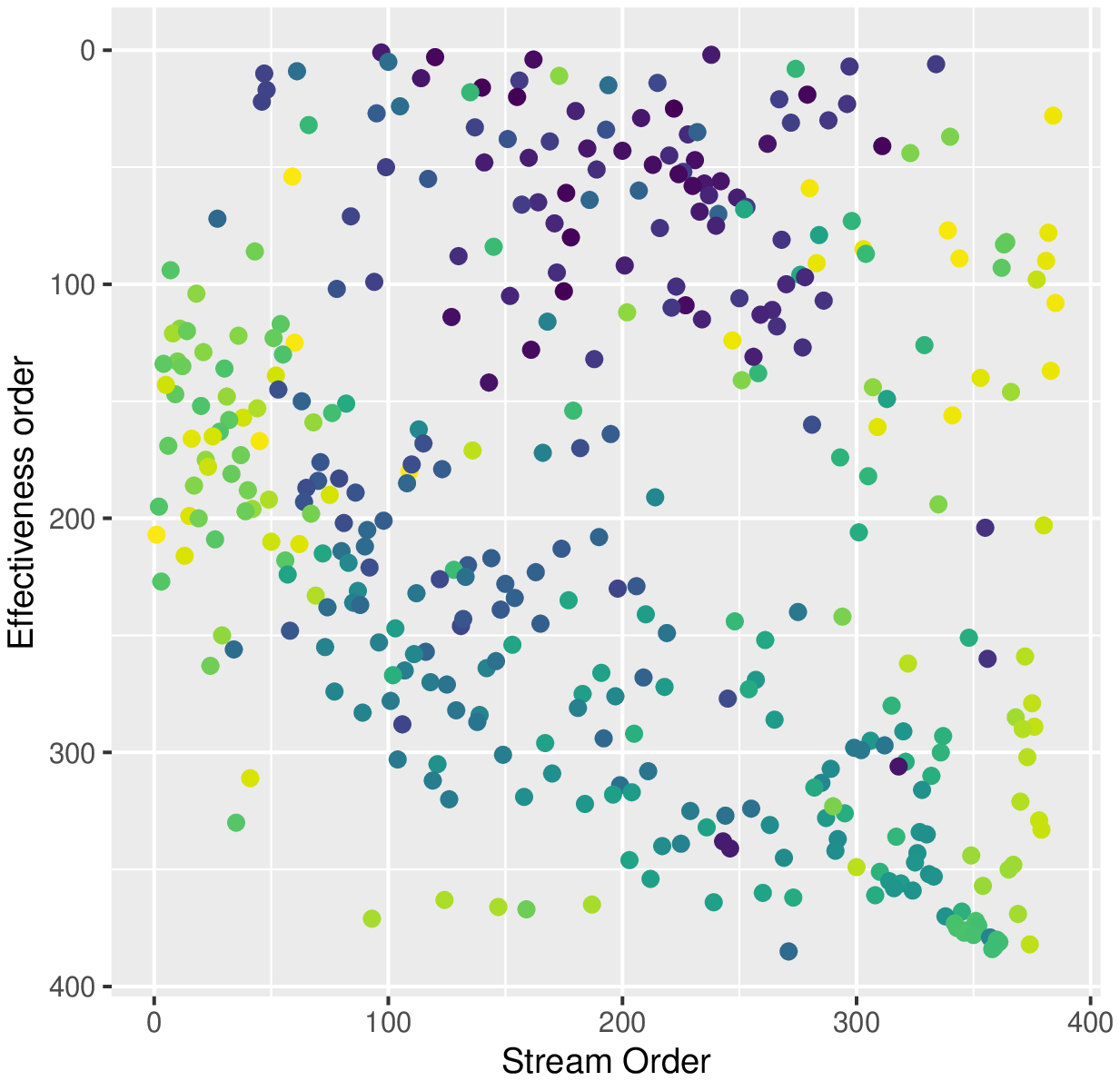}
    \caption{Correspondences between effectiveness order, stream order, and classification order.}
    \label{eight}
  \end{minipage}
\end{figure}

Figure \ref{nine} displays $\text{Effectiveness}(k)$ in a descending order (i.e., effectiveness order).
Figure \ref{eight} displays this order against the stream order, while the color indicates the classification order.
We observe two clusters in this figure.
Regarding the classification order, the cluster of sectors with the highest effectiveness (e.g., Coal mining, crude petroleum and natural gas, Metallic ores, Miscellaneous edible crops, etc.) are typically primary industries (with dark colors).
Another cluster is placed on the lower left-hand side of the figure.
Regarding the classification order, these are basically secondary and tertiary sectors. 
Regarding the stream order, upstream sectors have larger effectiveness and downstream sectors have smaller effectiveness.

\begin{figure}[t!]
\centering
   \begin{minipage}[b]{0.475\textwidth}
    \includegraphics[width=\textwidth]{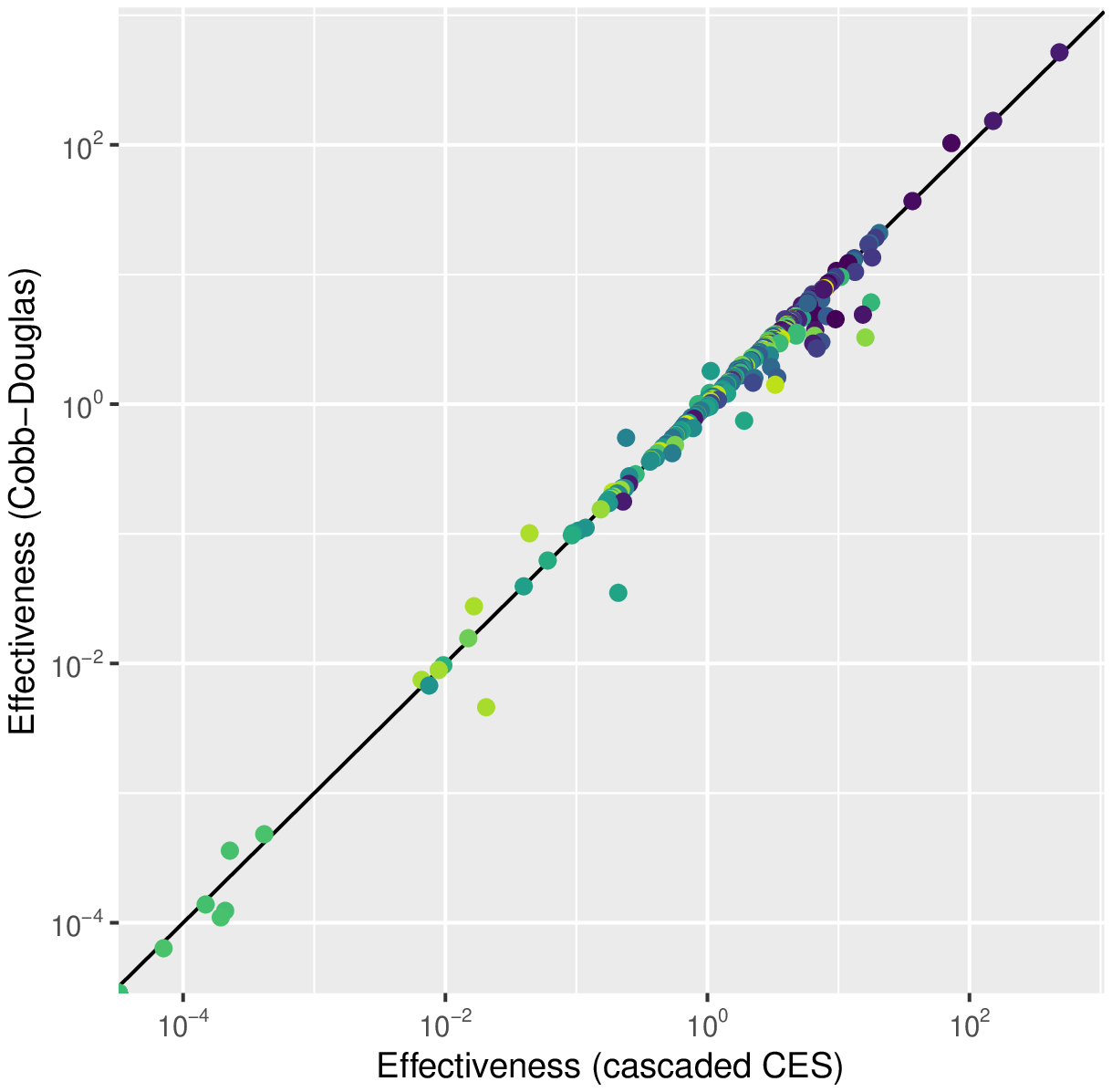}
    \caption{Correspondences between sectoral effectivenesses of empirical cascaded CES and Cobb--Douglas systems.}
    \label{nine2}
  \end{minipage}
\hspace{0.03\textwidth}
  \begin{minipage}[b]{0.475\textwidth}
    \includegraphics[width=\textwidth]{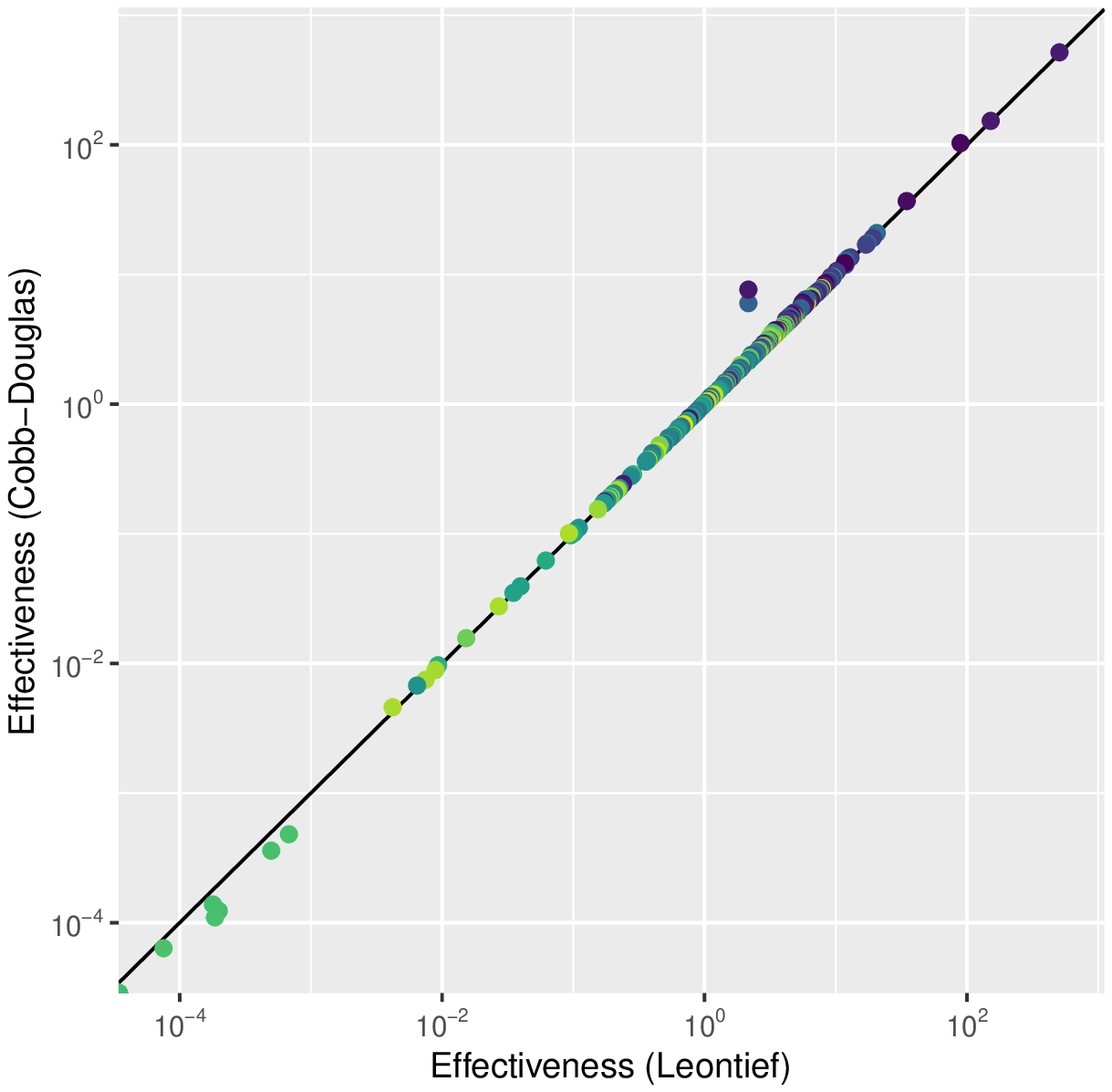}
    \caption{Correspondences between sectoral effectivenesses of Leontief and Cobb--Douglas systems.}
    \label{eight2}
  \end{minipage}
\end{figure}
In Figure \ref{nine2} we make comparison between sectoral effectivenesses of empirical cascaded CES and uniform unit elasticity (i.e., Cobb--Douglas) systems.
Note that Cobb--Douglas system can be obtained by setting $1-\gamma_{nj}=1$ for all $n$ and $j$, via the cascaded CES system.
Similarly, Leontief system can be obtained by setting $1-\gamma_{nj}=0$.
Figure \ref{eight2} compares sectoral effectivenesses of Cobb--Douglas and Leontief systems.
The latent possibility of factor substitution does not necessarily have the enhancing effect on the overall performance (cost effectiveness) of the economy, let alone distributional modifications.
However, the empirical cascaded CES system attains more effectiveness overall in comparison to the Cobb--Douglas system which in turn attains slightly more effectiveness overall in comparison to the inflexible Leontief system.

\section{Concluding Remarks}
In this study, we essentially model the \textit{metastructure} of an economy that overarches structural transformations observed in a set of linked input--output tables.
In this model, a shorter yet detailed structural transformation of the metastructure can be studied.
As the model encloses potential alternative technologies for each sector, it comprises a set of sectoral production functions spanning many substitutable factor inputs.
Each sectoral production is modeled by binary compounding processes ultimately cascaded under a universal sequence.
As we discover a self-similar hierarchical structure stylized in the empirical input--output transactions, we utilize a corresponding sequence as the fundamental and persistent structure behind structural transformations.

For all sectoral productions, by assuming constant returns to scale, we find that CES elasticity and share parameters for all binary compounding processes can be estimated by means of nest-wise dynamic programming, using factor-wise cost-share observations.
Moreover, sectoral TFP can be measured by the gap between the output of the cascaded CES and the observed output.
If parameters are estimated with two-point regressions from the corresponding observations, the two temporary distant cost-share structures can be completely restored by the feedback system of empirical dual functions of cascaded CES, together with the sectoral growths of TFP.

In addition, we model the utility of a representative household by multifactor CES function with single substitution elasticity. 
Our approach concerning households' expenditure shares enables us to estimate substitution elasticity using a fixed-effects regression that exploits the quantitative variety of the commodities consumed.
Regarding our data, the shape of the representative utility is essentially Leontief.
We then integrate the corresponding indirect utility function with a system of state-restoring unit cost functions, creating a dynamic general equilibrium model that evaluates welfare changes associated with household expenditures and labor demand by extrapolating structural transformations.

For demonstration, we independently inject marginal productivity shocks of positive and equal monetary value into each sectoral production and evaluate the resulting welfare changes in light of structural transformations. 
Although the magnitude differs significantly, this experiment results in positive welfare gains in all cases.
Regarding order of magnitudes of gainable welfare, primary sectors (i.e., farming, fishing, and mining) are effective overall.
We discover that upstream sectors are more effective than downstream ones with regard to secondary (i.e., manufacturing and processing) and tertiary (i.e., service) sectors. 
Cascaded CES has a lot of scope for modification although its flexibility and versatility should lead to a variety of applications concerning the modeling of the economy's metastructure.

\clearpage
\section*{\hypertarget{app}{Appendix}: Substitution Elasticity of Cascaded CES Functions}
We first examine the elasticities of a cascaded (serially nested) function between different factor inputs.
We begin by taking the partial derivative of (\ref{ccesucf}) with respect to $p_{i}$ and $p_{j}$ where we assume that ${n}>{i}>{j}$:
\begin{align}
\frac{\partial C}{\partial p_{i}}
&=\frac{\partial C}{\partial \pi_{N}} \cdots 
\frac{\partial \pi_{n+1} }{\partial \pi_{n}} 
\frac{\partial \pi_{n} }{\partial \pi_{n-1}} \cdots
\frac{\partial \pi_{i+1} }{\partial p_{i}} \\
\frac{\partial C}{\partial p_{j}}
&=\frac{\partial C}{\partial \pi_{N}} \cdots 
\frac{\partial \pi_{n+1} }{\partial \pi_{n}} 
\frac{\partial \pi_{n} }{\partial \pi_{n-1}} \cdots
\frac{\partial \pi_{i+1} }{\partial \pi_{i}}
\frac{\partial \pi_{i} }{\partial \pi_{i-1}} \cdots  
\frac{\partial \pi_{j+1} }{\partial p_{j}}
\end{align}
Further differentiating partially by $p_n$ gives the following:
\begin{align}
\frac{\partial^2 C}{\partial p_i \partial p_n}
&=\frac{\partial}{\partial p_n}\left( \frac{\partial C}{\partial \pi_{N}} \cdots \frac{\partial \pi_{n+1} }{\partial \pi_{n}} \right)
\frac{\partial \pi_{n} }{\partial \pi_{n-1}} \cdots 
\frac{\partial \pi_{i+1} }{\partial p_{i}}\\
\frac{\partial^2 C}{\partial p_j \partial p_n}
&=\frac{\partial}{\partial p_n}\left( \frac{\partial C}{\partial \pi_{N}} \cdots \frac{\partial \pi_{n+1} }{\partial \pi_{n}} \right)
\frac{\partial \pi_{n} }{\partial \pi_{n-1}} \cdots 
\frac{\partial \pi_{i+1} }{\partial \pi_{i}}
\frac{\partial \pi_{k} }{\partial \pi_{i-1}} \cdots  
\frac{\partial \pi_{j+1} }{\partial p_{j}}
\end{align}
Then, we find that:
\begin{align}
\frac{\frac{\partial^2 C}{\partial p_i \partial p_n}}{\frac{\partial C}{\partial p_{i}}}
=
\frac{\frac{\partial^2 C}{\partial p_j \partial p_n}}{\frac{\partial C}{\partial p_{j}}}
=
\left( \frac{\partial C }{\partial \pi_{N}} \cdots \frac{\partial \pi_{n+1} }{\partial \pi_{n}} \right)^{-1}
\frac{\partial}{\partial p_n}\left( \frac{\partial C_{N} }{\partial \pi_{N}} \cdots \frac{\partial \pi_{n+1} }{\partial \pi_{n}} \right)
\end{align}
The above argument depends only on the nest $n$ as long as the paired factor's nest is inside (i.e., $n>i, j$).
Hence, the Allen--Uzawa elasticity of substitution (AUES, denoted by $\eta^{\text{AU}}$) between nest $n$ and any nest inside of $n$, such as $i$ and $j$, will depend only on $n$. 
That is,
\begin{align}
\frac{C}{\frac{\partial C}{\partial p_{n}}}\frac{\frac{\partial^2 C}{\partial p_i \partial p_n}}{\frac{\partial C}{\partial p_{i}}}
=\eta_{ni}^{\text{AU}}
=
\frac{C}{\frac{\partial C}{\partial p_{n}}}\frac{\frac{\partial^2 C}{\partial p_j \partial p_n}}{\frac{\partial C}{\partial p_{j}}}
=\eta_{nj}^{\text{AU}}
=\eta_n^{\text{AU}}
&&
 n>i, j
\label{aues}
\end{align}
In other words, the AUES between an input and its inner-nest inputs are the same, while those between an input and its outer-nest inputs are not necessarily the same.
In Table \ref{tab_aues}, we highlight the same AUES with the same tone for the four-nest five-input case.

While AUES is a multifactor generalization of the two-factor elasticity of substitution, Morishima's elasticity of substitution (or MES, denoted by $\eta^{\text{M}}$) is a multi-factor generalization of the original elasticity of substitution concept.\footnote{Characteristic relations between Allen--Uzawa and Morishima elasticities of substitution are discussed in detail in \citet{standup}. }
MES can be defined via AUES as follows:
\begin{align}
\eta_{ij}^{\text{M}}=a_j \left(\eta_{ij}^{\text{AU}}-\eta_{jj}^{\text{AU}} \right)
\end{align}
Here, $a_j$ indicates the $j$th factor's cost share.
Note that while AUES is symmetrical (i.e., $\eta_{ij}^{\text{AU}}=\eta_{ji}^{\text{AU}}$ for any $i\neq j$), MES is not necessarily so.
Hence, with regard to (\ref{aues}), AUES is the same for all inner nest inputs relative to the reference nest input.
That is, if $i>j$, then $\eta_{ij}^{\text{AU}}=\eta_{i}^{\text{AU}}$, while if $i<j$, then $\eta_{ij}^{\text{AU}}=\eta_{j}^{\text{AU}}$.
This leads to the following exposition of MES for a cascaded function:
\begin{align}
\begin{aligned}
\eta_{ij}^{\text{M}}&=a_j \left(\eta_{ij}^{\text{AU}}-\eta_{jj}^{\text{AU}} \right)
=a_j \left( \eta_{i}^{\text{AU}}-\eta_{jj}^{\text{AU}}  \right) 
&&
i>j   
\\
\eta_{ij}^{\text{M}}&=a_j \left(\eta_{ij}^{\text{AU}}-\eta_{jj}^{\text{AU}} \right)
=a_j \left( \eta_{j}^{\text{AU}}-\eta_{jj}^{\text{AU}}  \right) 
= \eta_j^{\text{M}}
&&
j>i
\end{aligned} 
\label{moes}
\end{align}
In Table \ref{tab_moes} we highlight the same MES with the same tone for the four-nest five-input case.

Below, we examine the elasticities of a cascaded CES function of $N+1$ factor inputs.
Without loss of generality, we focus on the $n$th nest and write the unit cost function as follows:
\begin{align}
\pi = C\left( W_{n} \right) 
&&
W_n =\left(\pi_{n+1}\right)^{\gamma_{n}} 
= \alpha_{n} \left(p_{n}\right)^{\gamma_{n}} + \left( 1 - \alpha_{n} \right)\left(\pi_{n}\right)^{\gamma_{n}} 
\label{hq}
\end{align}
where ${n} = 1,\cdots, N$, and $\pi_1 = p_0$.
Note that $W_n$ is the $n$th compound price, raised to the power of $\gamma_{n}$ and that the remaining variables are included in $C$.
We hereafter use $C^\prime = \frac{\diff C}{\diff W_n}$ and $C^{\prime\prime}=\frac{\diff^2C}{\diff (W_n)^2}$.
For later convenience, we note that at the last nest, ${n} =N$, we must have
\begin{align}
C\left( W_N \right)=(W_N)^{{1}/{\gamma_N}} 
=\pi_{N+1}
\label{lastn}
\end{align}

The key partial derivatives for examining the elasticities between nest inputs ${n}$ and ${n}-1$ 
follow below:
\begin{align}
\frac{\partial C}{\partial p_{n}}
&= 
C^\prime \alpha_{n}\gamma_{n} (p_{n})^{\gamma_{n}-1} 
\\
\frac{\partial C}{\partial p_{{n}-1}}
&= C^{\prime} \alpha_{{n}-1} \left( 1 - \alpha_{n}\right) \gamma_{n} (p_{{n}-1})^{\gamma_{{n}-1} - 1} (\pi_{n})^{\gamma_{n} - \gamma_{{n}-1}}
\\
\frac{\partial^2 C}{\partial p_{n} \partial p_{{n}-1}}
&= 
C^{\prime \prime} \alpha_{n}\alpha_{{n}-1} \left( 1 - \alpha_{n}\right) (\gamma_{n})^2 (p_{n})^{\gamma_{n} - 1}(p_{{n}-1})^{\gamma_{{n}-1} - 1} (\pi_{n})^{\gamma_{n} - \gamma_{{n}-1}}
\\
\frac{\partial^2 C}{\partial p_{n}^2}
&=\alpha_{n}\gamma_{n} (p_{n})^{\gamma_{n} - 1} \left( 
C^{\prime\prime}
\alpha_{n} \gamma_{n} (p_{n})^{\gamma_{n} -1}  + C^{\prime} \left(\gamma_{n} -1 \right) (p_{n})^{-1} \right) 
\end{align}
The AUES of ${n}-1$ with respect to ${n}$ for a cascaded CES function can thus be evaluated as follows:
\begin{align}
\eta_{{n}-1 \, {n}}^{\text{AU}} 
=
\frac{C}{\frac{\partial C}{\partial p_{n}}}\frac{\frac{\partial^2 C}{\partial p_{n} \partial p_{{n}-1}}}{\frac{\partial C}{\partial p_{{n}-1}}}
= \frac{C}{C^{\prime}}\frac{C^{\prime \prime}}{C^{\prime}}
\label{bbaues}
\end{align}
Hence, the AUES for a cascaded CES function can vary depending on the $i$th and subsequent inner-factor prices.
However, an exception is the last nest, where (\ref{lastn}) has the following exposition:
\begin{align}
\eta_{N-1 \, N}^{\text{AU}} 
= \frac{C}{C^{\prime}}\frac{C^{\prime \prime}}{C^{\prime}}
= \frac{(W_N)^{1/\gamma_N}}{\frac{(W_N)^{-1+1/\gamma_N}}{\gamma_N}}\frac{\frac{\left( \frac{1-\gamma_N}{\gamma_N} \right)(W_N)^{-2+1/\gamma_N}}{\gamma_N}}{\frac{(W_N)^{-1+1/\gamma_N}}{\gamma_N}}
= 1 - \gamma_N
\label{auesn}
\end{align}

\begin{table}[thpb!]
\begin{center}
\begin{tabular}{c}
\begin{minipage}{0.5\hsize}
\begin{center}
\caption{AUES of a cascaded CES function ($N=4$).}
\label{tab_aues}
\begin{tabularx}{0.9\textwidth}{|c|CCCCC|}
\hline
& $4$ & $3$ & $2$ & $1$ & $0$  \\
\hline
$4$ & $-$ 
& \cellcolor{orange!20}{$1- \gamma_4$}
& \cellcolor{orange!20}{$1- \gamma_4$}
& \cellcolor{orange!20}{$1- \gamma_4$}
& \cellcolor{orange!20}{$1- \gamma_4$}
\\
$3$ & \cellcolor{orange!20}{$1- \gamma_4$} & $-$ 
& \cellcolor{orange!45}{$\eta_{3}$}
& \cellcolor{orange!45}{$\eta_{3} $}
& \cellcolor{orange!45}{$\eta_{3}$}
\\
$2$ & \cellcolor{orange!20}{$1- \gamma_4$} & \cellcolor{orange!45}{$\eta_{3}$} & $-$ 
& \cellcolor{orange!70}{$\eta_{2}$}
& \cellcolor{orange!70}{$\eta_{2}$}
\\
$1$ & \cellcolor{orange!20}{$1- \gamma_4$} & \cellcolor{orange!45}{$\eta_{3}$} & \cellcolor{orange!70}{$\eta_{2}$} & $-$ 
& \cellcolor{orange!95}{$\eta_{1}$}
\\
$0$ & \cellcolor{orange!20}{$1- \gamma_4$} & \cellcolor{orange!45}{$\eta_{3}$} & \cellcolor{orange!70}{$\eta_{2}$} & \cellcolor{orange!95}{$\eta_{1}$} & $-$\\\hline
\end{tabularx}
\end{center}
\end{minipage}
\begin{minipage}{0.5\hsize}
\begin{center}
\caption{MES of a cascaded CES function ($N=4$).}
\label{tab_moes}
\begin{tabularx}{0.9\textwidth}{|c|CCCCC|}
\hline
& $4$ & 3 & 2 & 1 & 0  \\
\hline
$4$ & $-$ & {$\eta_{43} $} & {$\eta_{42}$} & {$\eta_{41}$} & {$\eta_{40}$}\\
3 & \cellcolor{olive!20}{$1-\gamma_4$} & $-$ & {$\eta_{32}$} & {$\eta_{31}$} & {$\eta_{30}$} \\
2 & \cellcolor{olive!20}{$1-\gamma_4$} & \cellcolor{olive!40}{$1-\gamma_3$} & $-$ & {$\eta_{21}$} & {$\eta_{20}$}\\
1 & \cellcolor{olive!20}{$1-\gamma_4$} & \cellcolor{olive!45}{$1-\gamma_3$} & \cellcolor{olive!60}{$1-\gamma_2$} & $-$ 
& {$1-\gamma_1$}
\\
0 & \cellcolor{olive!20}{$1-\gamma_4$} & \cellcolor{olive!40}{$1-\gamma_3$} & \cellcolor{olive!60}{$1-\gamma_2$} & \cellcolor{olive!80}{$1-\gamma_1$} & $-$\\\hline
\end{tabularx}
\end{center}
\end{minipage}
\end{tabular}
\end{center}
\end{table}

In Table \ref{tab_aues}, we summarize AUES with regard to (\ref{aues}) and (\ref{auesn}).
The elasticities, which are symmetrical and equal among the inputs nested inside, equal the last parameter $1-\gamma_N$ when elasticities are evaluated with respect to the last input.
The MES of ${n}-1$ with respect to ${n}$ can be evaluated in the same manner.
\begin{align}
\eta_{{n}-1 \, {n}}^{\text{M}}  &= a_{n} \left( \eta^{\text{AU}}_{{n}-1 \, {n}} - \eta^{\text{AU}}_{nn} \right)  
=\frac{\partial C}{\partial p_{n}}\frac{p_{n}}{C}
\left( 
\frac{C}{\frac{\partial C}{\partial p_{n}}}\frac{\frac{\partial^2 C}{\partial p_{n} \partial p_{{n}-1}}}{\frac{\partial C}{\partial p_{{n}-1}}} 
- \frac{C}{\frac{\partial C}{\partial p_{n}}}\frac{\frac{\partial^2 C}{\partial p_{n}^2}}{\frac{\partial C}{\partial p_{n}}}
\right)
={p_{n}} 
\left( 
\frac{\frac{\partial^2 C}{\partial p_{n} \partial p_{{n}-1}}}{\frac{\partial C}{\partial p_{{n}-1}}} 
- 
\frac{\frac{\partial^2 C}{\partial p_{n}^2}}{\frac{\partial C}{\partial p_{n}}} \right)  
\\
&=p_{n}\left( 
\frac{
C^{\prime\prime}
\alpha_{n} \gamma_{n} (p_{n})^{\gamma_{n} - 1}}{
C^{\prime}
}
-
\frac{
C^{\prime\prime}
\alpha_{n} \gamma_{n} (p_{n})^{\gamma_{n} - 1}  +
C^{\prime}
\left( \gamma_{n} - 1 \right) (p_{n})^{-1}  }{
C^{\prime}
}
\right)
=1- \gamma_{n}
\label{moesn}
\end{align}
Regarding (\ref{moesn}), the MES of a nest input with respect to a back-to-back inner nest input is constant at the CES elasticity parameter of that nest.  
At the same time, according to (\ref{moes}), a nest input MES is the same with respect to any inner nest input.
Hence, a nest input MES with respect to any inner nest input is constant at the CES elasticity parameter of that nest.
In Table \ref{tab_moes} we summarize MES for a cascaded CES function.

Finally, we show that $\eta^{\text{M}}_{10}=\eta^{\text{M}}_{01}=1-\gamma_1$.
Below is a list of the partial derivatives we use to assess the MES for the nest at the core, i.e., $n=1$:
\begin{align}
\frac{\partial C}{\partial p_1}
&= C^\prime \alpha_{1}\gamma_1 (p_1)^{\gamma_{1}-1} 
\\
\frac{\partial C}{\partial p_{0}}
&= C^{\prime} \left( 1 - \alpha_{1}\right) \gamma_{1} (p_{0})^{\gamma_{1} - 1} 
\\
\frac{\partial^2 C}{\partial p_1 \partial p_{0}}
&= C^{\prime \prime} \alpha _{1} \left( 1 - \alpha_{1}\right) (\gamma_{1})^2 (p_{1})^{\gamma_1 - 1}(p_{0})^{\gamma_{1} - 1} 
\\
\frac{\partial^2 C}{\partial p_0^2}
&=\left(1 - \alpha_1\right) \gamma_1 (p_0)^{\gamma_1 - 1} 
\left( C^{\prime\prime} \left(1 - \alpha_1\right) \gamma_1 (p_0)^{\gamma_1 - 1}
+ C^{\prime} \left(\gamma_1 -1 \right) (p_0)^{-1} \right) 
\end{align}
Using the above terms, we acquire the following:
\begin{align}
\eta^{\text{M}}_{10}
&=\frac{\partial C}{\partial p_0}\frac{p_0}{C}
\left( 
\frac{C}{\frac{\partial C}{\partial p_{0}}}\frac{\frac{\partial^2 C}{\partial p_{0} \partial p_{1}}}{\frac{\partial C}{\partial p_{1}}} 
- \frac{C}{\frac{\partial C}{\partial p_{0}}}\frac{\frac{\partial^2 C}{\partial p_{0}^2}}{\frac{\partial C}{\partial p_{0}}}
\right)
={p_{0}} 
\left( 
\frac{\frac{\partial^2 C}{\partial p_{0} \partial p_{1}}}{\frac{\partial C}{\partial p_{1}}} 
- 
\frac{\frac{\partial^2 C}{\partial p_{0}^2}}{\frac{\partial C}{\partial p_{0}}} \right) 
\\
&=p_{0}\left(\frac{C^{\prime \prime} \left( 1 - \alpha_{1}\right) \gamma_{1} (p_{0})^{\gamma_{1} - 1}}{C^{\prime}}-\frac{C^{\prime\prime} \left(1 - \alpha_1\right) \gamma_1 (p_0)^{\gamma_1 - 1}
+ C^{\prime} \left(\gamma_1 -1 \right) (p_0)^{-1}  }{C^\prime}\right)
=1-\gamma_1
\end{align}

\clearpage
{\raggedright
\bibliography{bibNN}
}
\end{document}